\documentclass[conference]{IEEEtran}
\usepackage{biblatex} 
\usepackage{geometry}                		% See geometry.pdf to learn the layout options. There are lots.
\usepackage[parfill]{parskip}    		        % Activate to begin paragraphs with an empty line rather than an indent
\usepackage{graphicx}				% Use pdf, png, jpg, or eps§ with pdflatex; use eps in DVI mode
\usepackage{amssymb}				% !TEX TS-program = latex   pdflatexmk
\usepackage{amsmath}
\usepackage{textcomp}
\usepackage{bbm}
\usepackage{algorithmic}
\usepackage[linesnumbered,ruled,vlined]{algorithm2e}
\usepackage{tabularx}
\usepackage{latexsym}
\usepackage{subcaption}
\usepackage{fancyhdr}
\usepackage{booktabs} 
\usepackage{eucal}
\usepackage{float}
\usepackage{parskip}
\usepackage{textcomp}
\usepackage{geometry}                		% See geometry.pdf to learn the layout options. There are lots.
\usepackage{tabularx}
\usepackage{mathrsfs}
\usepackage{calrsfs}
\usepackage{mathtools} 
\usepackage{stmaryrd}
\usepackage{wasysym} 
\usepackage{microtype}
\usepackage{yfonts}
\usepackage{listings} 
\usepackage{url}
\usepackage{multirow}
\usepackage{xcolor}
\usepackage{pdfpages}
\usepackage{graphicx} 
\usepackage{xurl}
\usepackage[utf8]{inputenc}
\usepackage [english]{babel}
\usepackage [autostyle, english = american]{csquotes}
\graphicspath{{./Images/}} 
\usepackage{mathtools}
\usepackage{amsmath}
\usepackage{fancyhdr} 
\usepackage{titlesec}

\usepackage[backref=true,colorlinks=true,linkcolor=blue, urlcolor=blue, citecolor=blue]{hyperref}
\hypersetup{
    pagebackref=true,
    colorlinks=true,
    citecolor=blue,
    linkcolor=blue,
    filecolor=magenta,      
    urlcolor=blue,
}

\definecolor{mygreen}{rgb}{0,0.6,0}
\definecolor{mygray}{rgb}{0.47,0.47,0.33}
\definecolor{myorange}{rgb}{0.8,0.4,0}
\definecolor{mywhite}{rgb}{0.98,0.98,0.98}
\definecolor{myblue}{rgb}{0.01,0.61,0.98}

\title{Embedded Microcontrol for Photovoltaic Water Pumping System}
\author{Justin London}

\titleformat{\chapter}[display]
{\normalfont\Huge\bfseries\raggedright}{\chaptertitlename\  7}
{15pt}{\Huge}
\begin{document} 

\author{\IEEEauthorblockN{Justin London \\}
\IEEEauthorblockA{\textit{Department of Electrical Engineering and Computer Science} \\
\textit{University of North Dakota}\\
Grand Forks, North Dakota USA \\
justin.london@und.edu}
}

\maketitle 

\begin{abstract}
    We introduce a novel 3-axis solar tracker water pumping system.  The charge generated from solar energy converted by the photovolatic panel (PV) cells is stored in a 12V battery that in turn powers two water diaphragm pumps using a solar charge controller that includes an MPPT algorithm that serves as a DC-DC converter.  The system is analyzed from an embedded microcontroller and embedded software perspective using Arduino. The photovoltaic panel uses four light photocell resistors (LPRs) which measure solar light intensity.  An ultrasonic sensor measures the water level in a reservoir water tank.  If the water level is too low, water is pumped from one water tank to the reservoir tank.  Using a soil moisture sensor, another water pump pumps water from the reservoir tank to the plant if water is needed.  Circuit designs for the system are provided as well as the embedded software used.  Simulation and experimental results are given.
\end{abstract}

\section{Introduction}
\ \ \ Photovoltaic systems (PVS) use solar irradiance and temperature sensor data to convert solar energy into electrical energy to power motors, and other actuator devices.  A PVS uses one or more photovoltaic cells to convert the sunlight into electrical energy through the photovoltaic effect process.  

The electrical charges generated by the cell is stored in batteries.  PVS are efficient renewable energy solutions to generate power for small and medium applications such for solar wind as they harness energy from the sun. Unlike fossil fuels, which are non-renewable and harm the environment with carbon emissions, PVs are clean. They are cost-effective to build especially in developing countries.  

As the cost of solar electricity has declined, \enquote{the number of grid-connected solar photovoltaics systems has grown into the millions and utility scale solar power stations with hundreds of megawatts are being built.}  Since the output power performance and efficiency of PVs is impacted by critical environmental factors such as temperature, humidity, dew point, cloud coverage, altitude, visibility, pressure, they must be considered in a PVs design.   In particular, the solar panel array must be designed for maximum sunlight exposure by facing perpendicular to the light waves that are incident to them.  The surface of the panels must not have dust on them.  Thus, an automatic real-time sun tracking system is needed for maximum possible efficiency or the panels will not generate their maximum output.

To illustrate the performance and efficiency of PV water pump systems, we introduce a novel 3-axis solar tracker water pumping system, in contrast to various dual axis-trackers that have been proposed \cite{Riley:2014}, \cite{Moron:2017}, that is applied to water plants and irrigation.  The solar tracker uses solar energy to power a 12 V DC lead acid battery that in turn powers two water pumps.  Water in one tank is pumped by one water pump to a second tank.  A second water pump pushes water from the second tank through a hose that can water plants via a spray nozzle.   We focus on the embedded systems used for control by using an Arduino microcontroller.  Arduino has been used in other PV solar energy and solar tracking research \cite{Jumaat:2018},\cite{Pramod:2019}, \cite{Khatib:2016}, \cite{Khatib:2021}.

The proposed system uses three sensors: (1) light photocell resistors (LPRs) (resistivity is a function of the incident electromagnetic radiation) to measure solar irradiance (light/photon intensity) from the sun; (2) ultrasonic sensor to measure the water level in a water tank; and (3) soil moisture sensor to measure the moisture in the soil of the plant.  An LCD display is used to show the water level and soil moisture measurements.

We focus on the embedded circuit design that is used to control the rotation, tilt, and movement of the PV panel to optimize conversion of solar energy to electrical through the photoelectric process to provide power the DC battery and the use of embedded software and algorithms to control two water diaphragm pumps.  We also analyze the use of pulse width modulation (PWM) to control the speed of the stepper motors and the water pumps.  

Arduino was selected as the microcontroller board due to its numerous sensor libraries available for integration and the board's ease of use for fast prototyping.  However, if implementation of the proposed system were for a commercial application, a PIC 16F877A microcontroller would be used instead as the PIC allows much more low level control at the pin and circuit level than the Arduino board and has more GPIO pins.

\subsection{MPPT}
Maximum power point tracking (MPPT) is an electronic DC to DC converter that optimizes the power match between the solar array (PV panels) and the battery (or battery bank). In essence, they convert and adjust a higher voltage DC output from solar panels down to the lower voltage needed to charge batteries.  All MPPTs are microprocessor controlled.   

A step-down (step-up) chopper or boost/DC-DC converter can be applied to MPPT systems in which the output voltage needs to be less (greater) than the input voltage. The relation between the output and input voltages ($V_{out}$ and $V_{in}$, respectively) is as follows
\begin{equation}
    \frac{V_{out}}{V_{in}} = \frac{1}{1-D}
\end{equation}
where $D$, $0 \leq D \leq 1$, is the duty cycle \cite{Amar:2022}.

Digital MPPT controllers know when to adjust the power being sent to the battery.  There are two common local optimizing hill climbing algorithms: (1) Perturb and Observe (P\&O) and (2) incremental conductance (IC).  
\subsection{Perturb and Observe}
\ \ P\&O determines on which side of the maximum power point (MPP), the hill climb optimization of the PV curve movies (left or right) by observing the sign of the previous change in voltage ($d$V) and change in power ($d$P). 

Depending on that information, the voltage increases (V++) or decreases (V--) the voltage.  If it is on the left side of the MPP, it moves right (increases) and if it is on the right of the MPP, the voltage moves left.  In particular, if the $d$V and $d$P are both positive or both negative, the voltage increases.  Otherwise, if dV and dP differ, the voltage decreases.  

The P\&O method uses periodic perturbation via incremental increases or decreases of voltage (or current) generated by a PV source, along with a trial assessment of the power output of the present cycle to the power output of the preceding perturbation cycle \cite{Sharma:2019}.  The P\&O algorithm continuously searches the MPP in subsequent perturbation cycles.   
%    Figure \ref{fig:mmpt} shows a flow diagram of the MMPT P\&O algorithm.
%\begin{figure}[H] 
%	\centering  \includegraphics[width=0.9\columnwidth]{Images/MMPT2.png} 
%	\caption{MMPT P\&O Flow Chart. Source: \cite{Sharma:2019}}
%	\label{fig:mmpt} 
%\end{figure} 
    Algorithm \ref{alg:mmpt2} shows the MMPT P\&O algorithm.
\begin{algorithm}
\caption{{Perturb and Observe Algorithm }}
\label{alg:mmpt2}
\textbf{Initialization:} $\tau = 0$,  $\mathcal{T}_{max}$, $\mathcal{I}(\tau)$, $\mathcal{V}(\tau)$\\
\If{$\tau \leq \mathcal{T}_{max}$}{
    $\mathcal{P}(\tau) = \mathcal{I}(\tau) \times \mathcal{V}(\tau)$ \\
    \eIf{$d\mathcal{P} > 0$} {
	   \eIf{$d\mathcal{V} > 0$} {
        {$\mathcal{V}_{ref} = \mathcal{V}_{ref} - d\mathcal{V}$} \\
      }{
        {$\mathcal{V}_{ref} = \mathcal{V}_{ref} + \mathcal{V}$} \\
      }
      \textbf{end} \\
      }
      {
        \eIf{$d\mathcal{V} < 0$} {
        {$\mathcal{V}_{ref} = \mathcal{V}_{ref} + d\mathcal{V}$} \\
      }{
        {$\mathcal{V}_{ref} = \mathcal{V}_{ref} - \mathcal{V}$} \\
      }
      \textbf{end} \\
      }
     \textbf{end} \\
%\textbf{end}
$\mathcal{I}(\tau)$ = $\mathcal{I}(\tau-1)$ \\
$\mathcal{V}(\tau)$ = $\mathcal{V}(\tau-1)$ \\
$\mathcal{P}(\tau)$ = $\mathcal{P}(\tau-1)$ \\
$\tau = \tau +1$ \\
}
\textbf{end}
\end{algorithm}

\subsection{Incremental Conductance}
    P\&O has the drawback that it oscillates around the MPP and may initially move in the incorrect direction when environmental conditions change.  Incremental conductance (IC) overcomes this issue is by representing the derivative of the power as:
    \begin{equation}
        \frac{dP}{dV} = \frac{d(I \cdot V)}{dV} \rightarrow \frac{1}{V}\frac{dP}{dV} = \frac{dI}{dV} + \frac{I}{V}
    \end{equation}
where $\frac{dI}{dV}$ is the incremental conductance.  The sign of the expression on the right side is always the same as the sign of the power change.
%\subsection{Q-Learning}
\section{Photovoltatic Array Model}
\begin{figure}[H] 
	\centering
	\includegraphics[width=0.85\columnwidth]{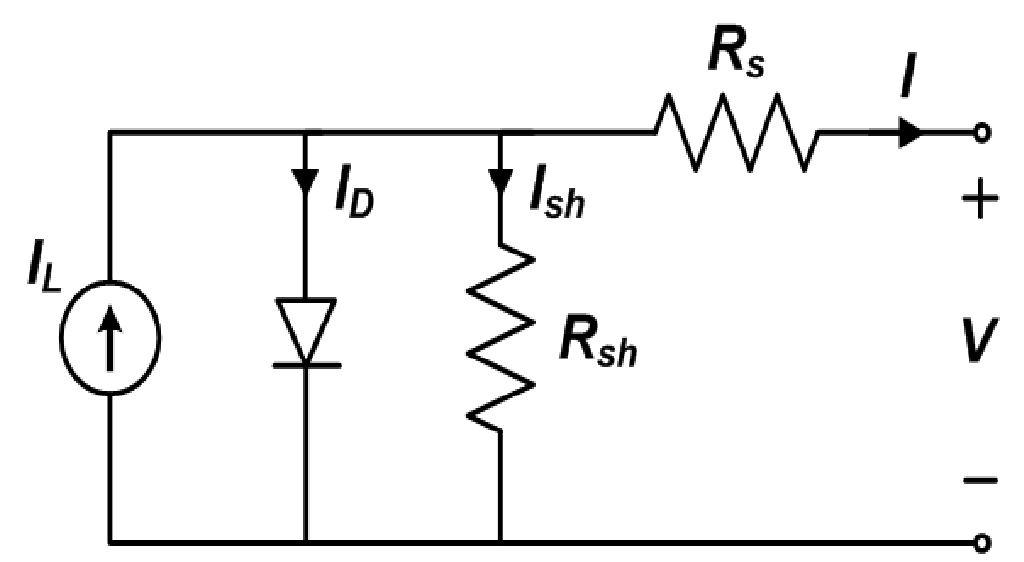} 
	\caption{PV cell as single diode circuit. Source: \cite{Sharma:2019}}
	\label{fig:cell} 
\end{figure}

\ \ \ A solar cell can be represented as a double-diode model.  The output current of a solar cell can be expressed as:
\begin{multline}
    I_{c} = I_{Ph} - I_{o_{1}} \bigg [ \text{exp} \bigg (\frac{V_{c}+I_{c}R_{s}}{V_{t_{1}}} \bigg )  -1 \bigg] \\ - I_{o_{2}} \bigg [\text{exp} \bigg (\frac{V_{c}+I_{c}R_{s}}{V_{t_{2}}} \bigg ) -1 \bigg]  - \frac{V_{c} + I_{c}R_{s}}{R_{p}}
\end{multline}
where $I_{c}$ and $V_{c}$ are the output current (A) and voltage (V) of the solar cell, respectively; $I_{ph}$ is the photocurrent; $I_{o_{1}}$ and $I_{o_{2}}$ are the diode saturation currents of the first and second diodes, respectively in ampere (A);  $R_{s}$ is the series resistance ($\Omega$); $R_{p}$ is the resistance ($\Omega$); and $V_{t_{1}}$ and $V_{t_{2}}$ are the diode thermal voltages given by:
\begin{equation} 
    V_{t_{1}} = \frac{a_{1}k_{B}T_{c}}{q}
\end{equation}
\begin{equation}
     V_{t_{2}} = \frac{a_{2}k_{B}T_{c}}{q}
\end{equation}
where $q$ is the electron charge (1.602 $\times 10^{-19}$ Colombs), $k_{B}$ is Boltzmann's constant (1.381 $\times 10^{-23}$ J/K), $T_{C}$ is the cell temperature (Kelvin), and $a_{1}$ and $a_{2}$ are the diode ideality factors that represent the components of the diffusion and recombination currents, respectively.  The PV array consists of $N_{s}$ and $N_{p}$ cell modules, which are connected in series and parallel, respectively, to constitute the power demand by the load.  The output current of a PV array can be expressed as:
    \begin{multline}
        I_{a} = N_{p}I_{Ph} - \\ N_{p}I_{o_{1}} \bigg [\text{exp} \bigg (\frac{1}{V_{t_{1}}} \bigg (\frac{V_{a}}{N_{s}} + \frac{I_{a}R_{s}}{N_{p}} \bigg ) \bigg )-1 \bigg] \\
        - N_{p}I_{o_{2}} \bigg [\text{exp} \bigg (\frac{1}{V_{t_{2}}} \bigg (\frac{V_{a}}{N_{s}} + \frac{I_{a}R_{s}}{N_{p}} \bigg ) \bigg ) -1 \bigg] \\ - \frac{N_{p}}{R_{p}} \bigg (\frac{V_{a}}{N_{s}} + \frac{I_{a}R_{s}}{N_{p}} \bigg )
    \end{multline}
    where $V_{a}$ and $I_{a}$ are the output voltage (V) and current (A) of the PV array, respectively.  The PV cell is essentially a p-n junction that is fabricated on a thin film semiconductor like silicon.  As the cell is exposed to light energy (photons), the photon that hits the cells will be absorbed by semiconducting material.  Solar panels are usually made of polycrystalline that run with 12V, 250 mA, 3W as a source.  In an ideal cell $R_{s}$ is 0 and $R_{sh}$ is infinite. The efficiency of the PV array can be computed as
    \begin{equation}
        \zeta_{PV}  =. \frac{V_{a}I_{a}}{AG_{T}}
    \end{equation}
    where $V_{a}$ and $I_{a}$ are the output voltage (V) and current (A) of the PV array, respectively.  Figure \ref{fig:cell} shows the equivalent circuit of a PV cell modeled as a single diode circuit.
\section{System Model}
\begin{figure}[H] 
	\centering
    \includegraphics[width=1\columnwidth]{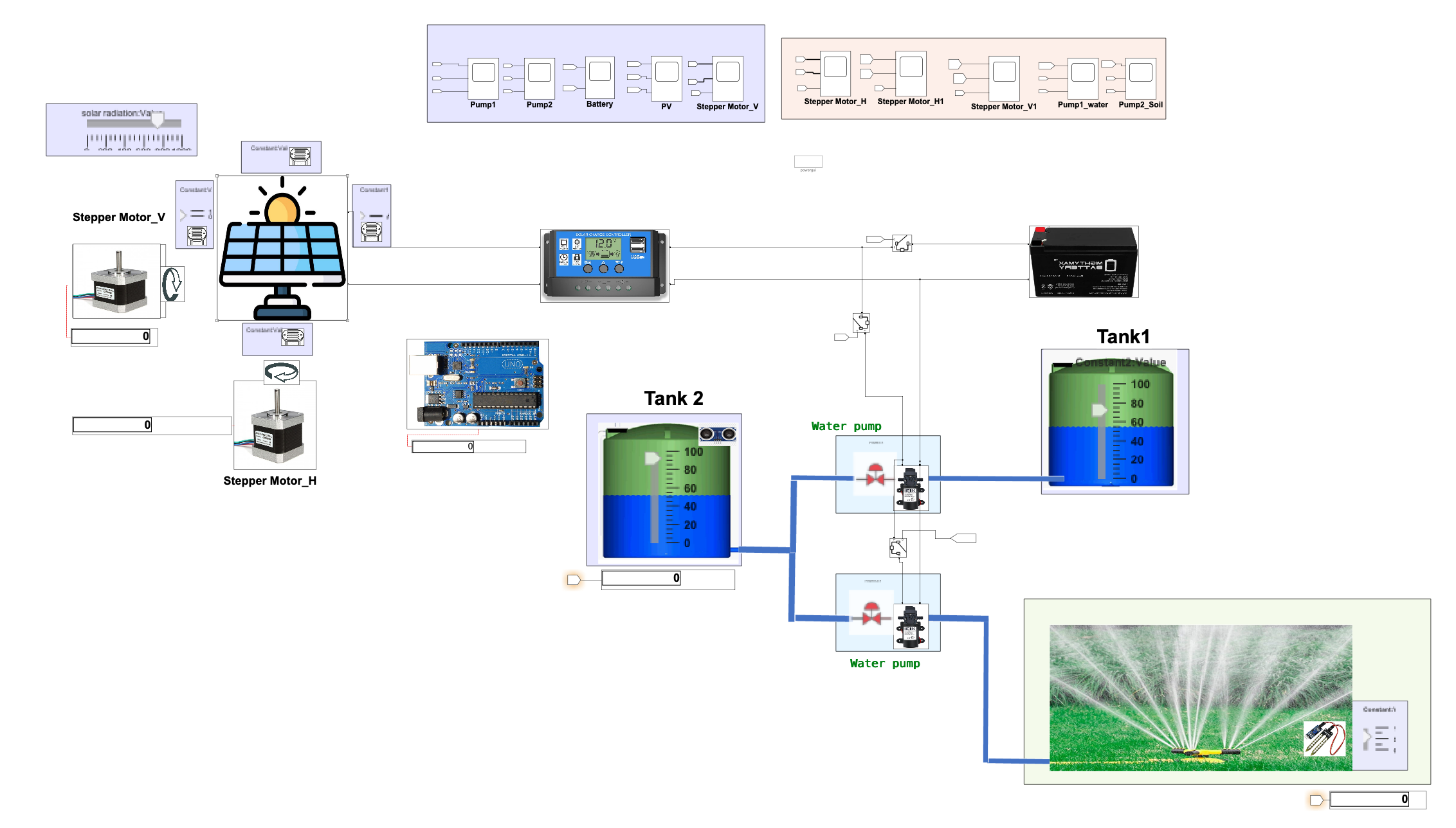} 
	\caption{PV Solar Tracking and Water Pumping Circuit Diagram}
	\label{fig:diagram3} 
\end{figure}

    The model system components are shown in Figure \ref{fig:diagram3}.  The solar management controller is connected to a 12V DC lead acid battery and charge relay switch to control the power provided to the water pump.    
    A water pump relay switch and a motorized ball valve are connected to each of the water pumps via a 2-way Y-valve connector. The water pumps are powered by the 12V DC lead acid battery that stores electric energy converted from solar energy via the photoelectric effect.  An ultrasonic sensor is placed under the lid of tank 2 and measures the water level. Each water tank is 13.8 liters.  If the water level is below $10\%$, the relay switch will turn on and water will be pumped from tank 1 into tank 2.  The switch will be turned off if the water level is $85\%$ full.  A soil moisture sensor measures the moisture level.  If the level is $< 500$, the soil is wet and should not be watered. Levels of $500-750$ indicate a humid target range.  A threshold level $> 750$ indicates dryness and therefore triggers the relay switch to turn on a water pump that pumps water through a hose and nozzle spray directed at the plant.  
    
    The DC motor model can be modeled using a PID controller and look-up tables.  The look-up table uses the PWM as an input to rotate the motor to a pre-determined angle. For instance, if the pulse width changes from 1.25 ms to 1.75 ms, the panel angle changes from $0^{\circ}$ to $180^{\circ}$ in a linear manner \cite{Chin:2011}. The second look-up table on the feedback path provides the actual pulse width results.  The actual and desired pulse-width are compared to obtain an error signal for a Proportional-Integral-Derivative (PID) controller to drive the motor to the desired angle \cite{Chin:2011}.
    
    Stepper motors are typically operated in open-loop systems due to their microstepping with discrete steps, but they can also used in closed-loop systems such as with encoder feedback.  The benefits of using feedback with stepper motors include not losing position when overloaded, handling higher torque loads as the feedback keeps the magnetic force optimally aligned, and the motor runs cooler sicne the feedback adjusts control current depending on the load.  The disadvantage is that they closed-loop stepper motors are more expensive than their open-loop counterparts.
\section{Circuit Design}
\ \ \   Figure \ref{fig:circuit} shows a schematic Simulink diagram of the PV solar tracking and water pumping system.
\begin{figure}[H] 
	\centering
\includegraphics[width=1\columnwidth]{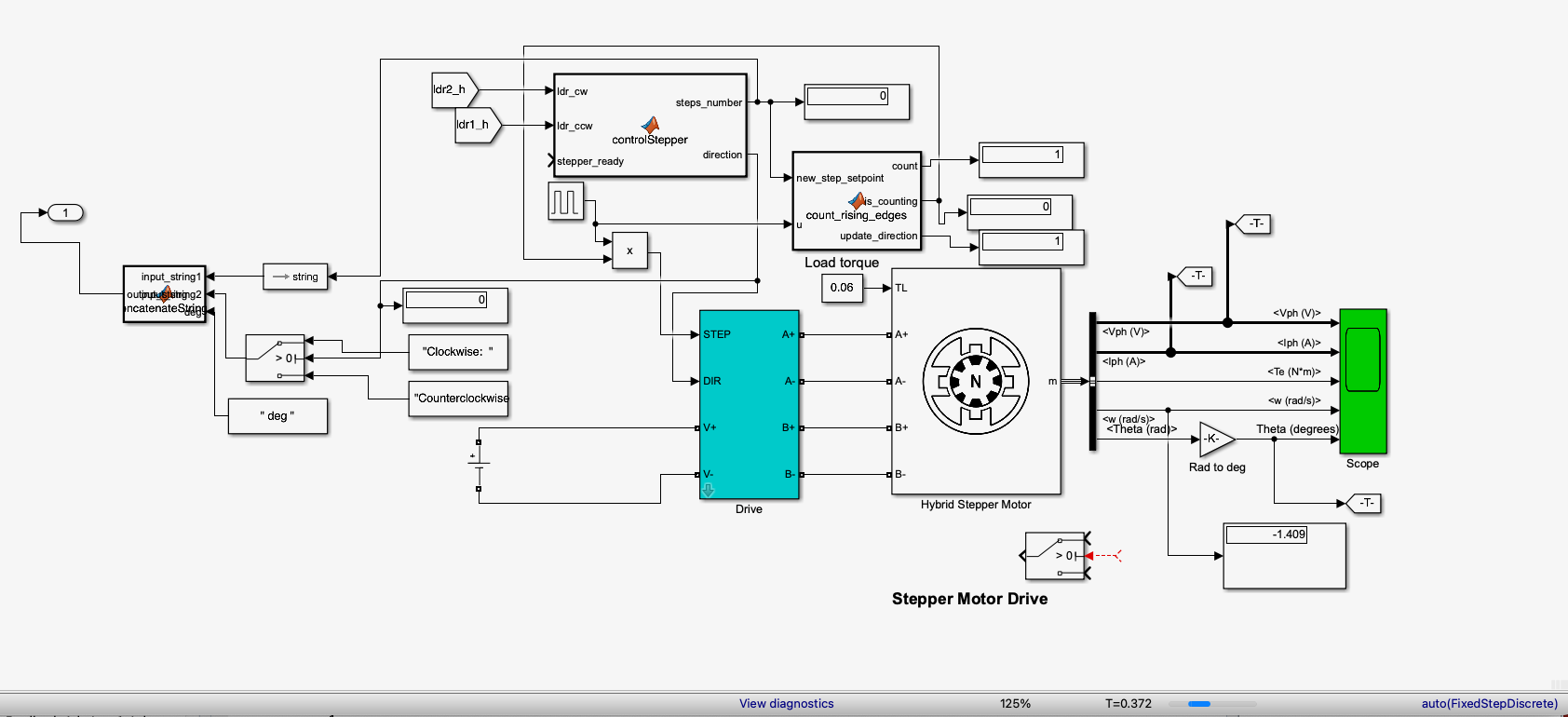} 
	\caption{PV Solar Tracking and Water Pumping Circuit Diagram}
	\label{fig:circuit} 
\end{figure}

\begin{figure}[H] 
	\centering
\includegraphics[width=1\columnwidth]{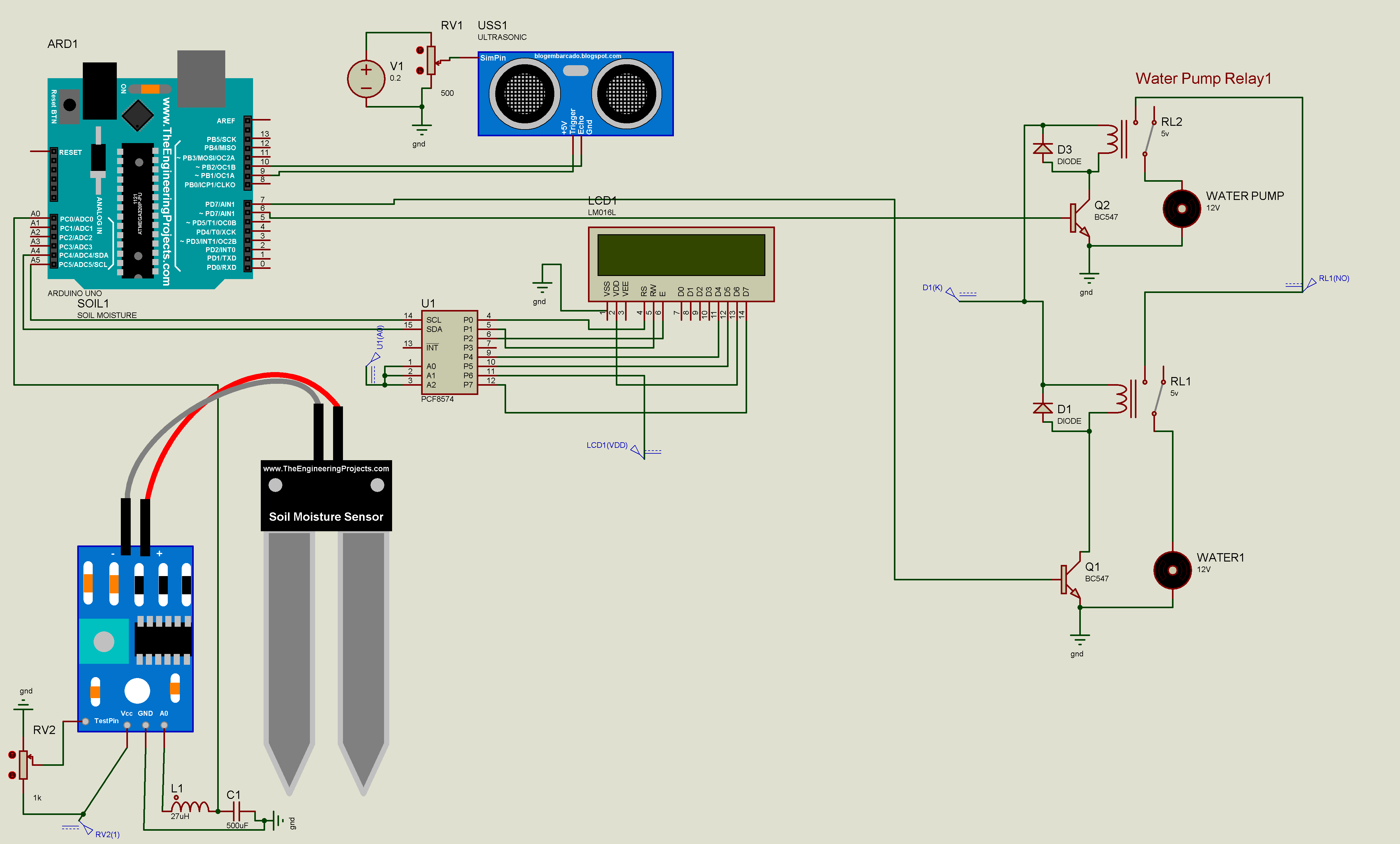} 
	\caption{Water Pumping Circuit Diagram}
	\label{fig:circuit3} 
\end{figure}

\begin{figure}[H] 
	\centering
\includegraphics[width=1\columnwidth]{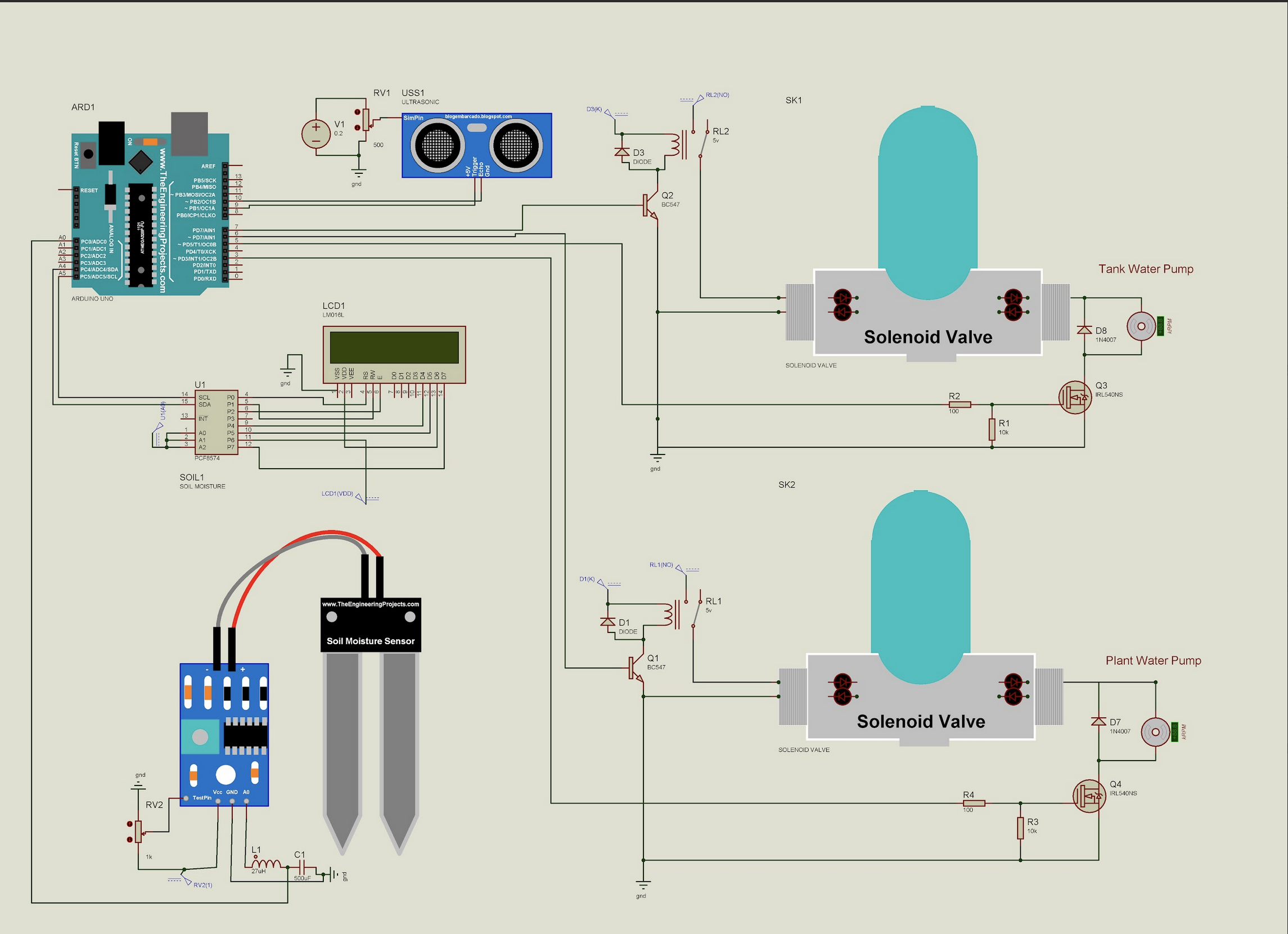} 
	\caption{Water Pumping Circuit Diagram with PWM}
	\label{fig:circuit4} 
\end{figure}

\begin{figure}[H] 
	\centering
\includegraphics[width=0.9\columnwidth]{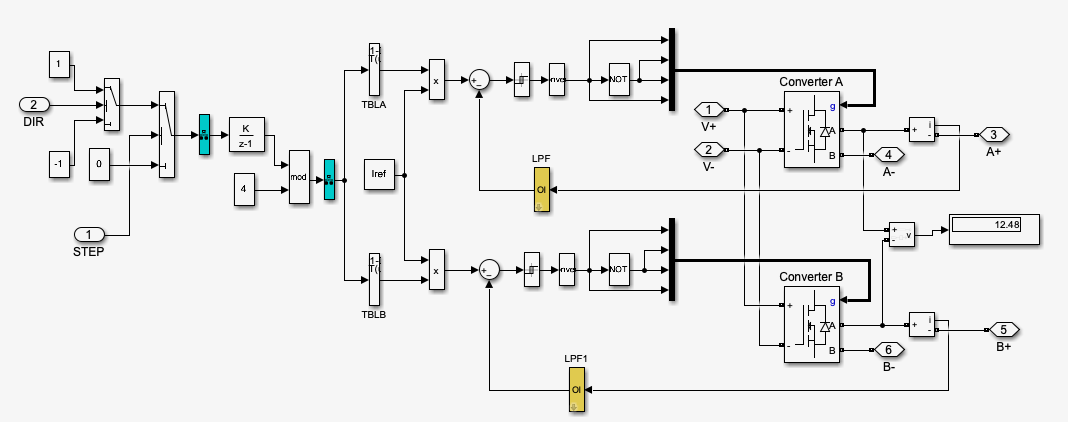} 
	\caption{DC-DC Converter and Driver Simulink Diagram}
	\label{fig:buck2} 
\end{figure}

\begin{figure}[H] 
	\centering
\includegraphics[width=0.9\columnwidth]{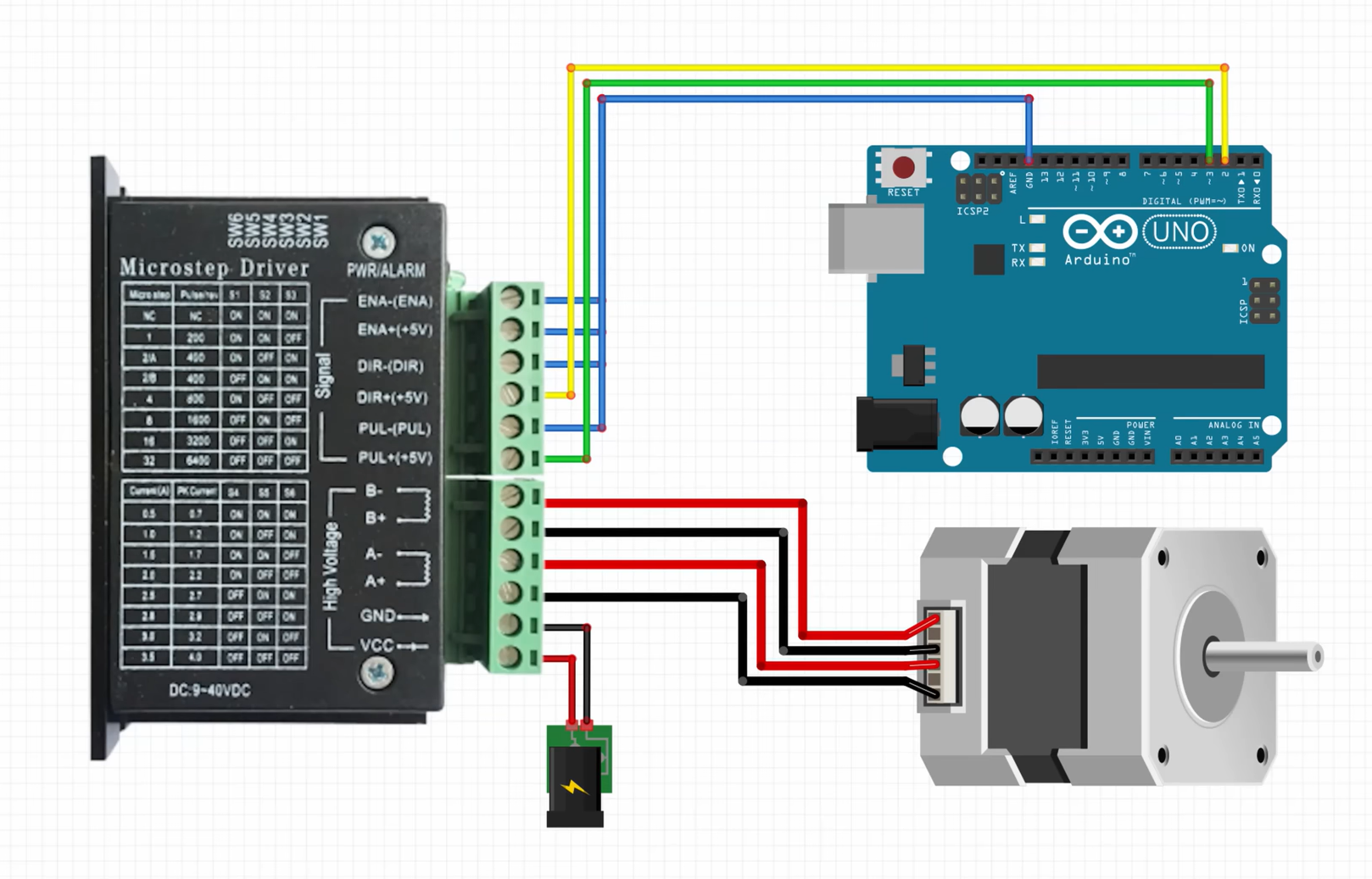} 
	\caption{Stepper Motor Wiring Diagram}
	\label{fig:circuit2} 
\end{figure}

\begin{figure}[H] 
	\centering
\includegraphics[width=0.9\columnwidth]{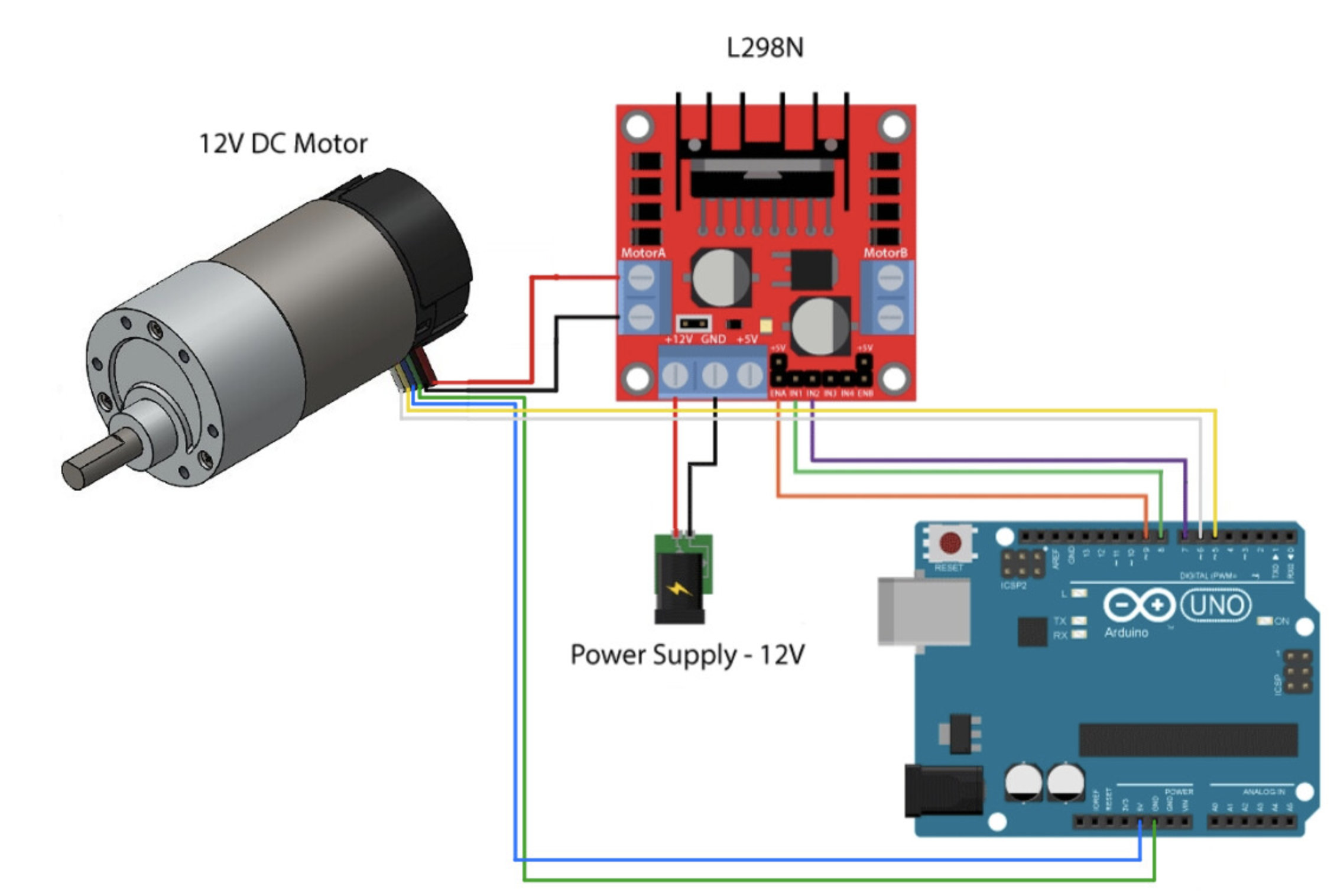} 
	\caption{DC Motor Wiring Diagram}
	\label{fig:dc2} 
\end{figure}

Figure \ref{fig:circuit} shows the detailed Proteus circuit diagram for the PV solar tracking and water pumping system.  The PV panel is 20 W.  It is connected to a solar management controller, which contains a built in DC-DC buck or step-down converter that controls the voltage.
\begin{figure}[H] 
	\centering
\includegraphics[width=1\columnwidth]{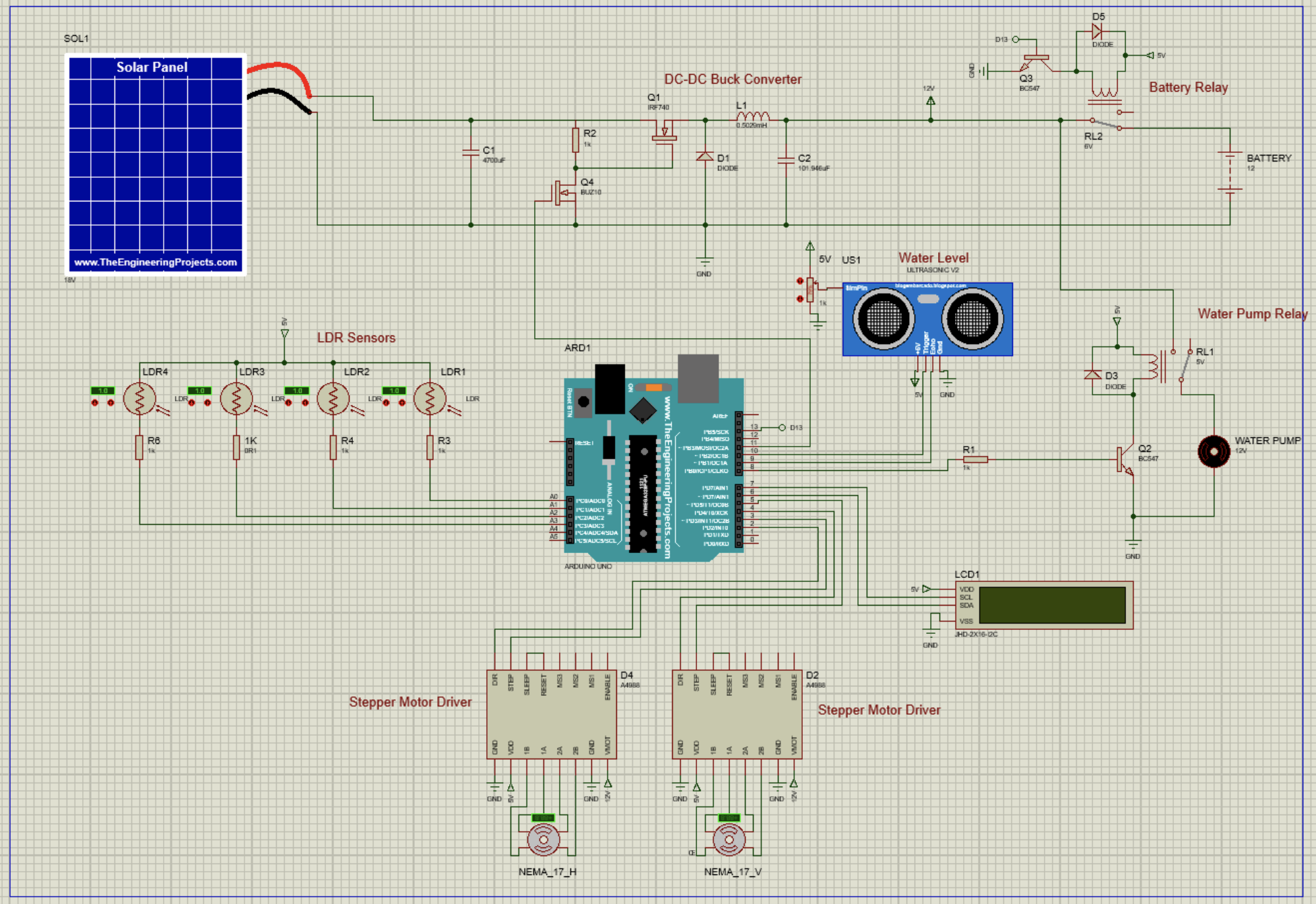} 
	\caption{PV Solar Tracking Circuit Diagram}
	\label{fig:circuit} 
\end{figure}
    Figure \ref{fig:circuit3} shows the circuit diagram for the water pumping subsystem which includes the relay switches and motorized ball valves connected to the water pump and the ultrasonic sensor to measure the water level.   
    Figure \ref{fig:circuit4} shows the above circuit modified to include PWM to control the speed of the motor in the water pumps.  This would be required in a commercial application where it is important to control the flow rate of water out of water pumps controlled by solenoid valves of a reservoir tank or deep well and into a plant tank for purification.
    The buck step-down converter decreases voltage, while increasing current, from its power source input to its (load) output.  Typically, the switch is a metal-oxide-semiconductor field-effect (MOSFET) or bipolar junction (BJT) transistor.    
\begin{figure}[H] 
	\centering
\includegraphics[width=0.9\columnwidth]{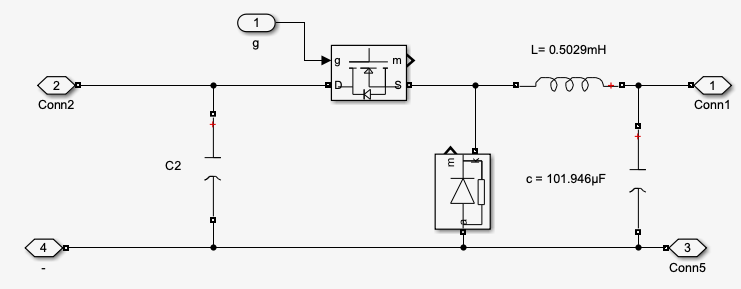} 
	\caption{DC-DC Converter Circuit Diagram}
	\label{fig:buck} 
\end{figure}
Figure \ref{fig:buck} shows the circuit diagram for the DC-DC converter. Figure \ref{fig:buck2} shows the Simulink DC-DC converter and motor driver connections. 
    The PV is a 3-axis solar tracker on a tripod that uses a Nema 23 stepper motor for rotation and a 12V DC motor for tilting based on the amount of solar irradiance and temperature.  A linear actuator raises and lower the tripod and panel.   The Nema 23 stepper motor is a bipolar two phase motor with and 200 steps/rev,  $1.8^{\circ}$ step resolution. Each phase (coil) draws current 2.8A (with a rate current of 600 mA) at 3.9 V, allowing for a holding torque of 1.26 Nm.  While a Nema 17 stepper was initially chosen, the overall weight of the PV frame along with the need for larger 8mm shaft necessitated using the Nema 23 instead.  The advantage of using the stepper motor is that it allows one to limit the   Figure \ref{fig:circuit2} shows the wiring diagram for a stepper motor.  The stepper motor is connected to and controlled by an %A4988 
    TB6600 stepper motor driver with a heat sink.   
    The gear 12V DC motor draws 1.6A of current (with a rated current of 350 mA), a rated speed of 12 rpm, and a rated torque of 70 kg/cm.  Figure \ref{fig:dc2} shows the wiring diagram for the 12V DC motor.

%Figure \ref{fig:buck_converter} shows a Simulink diagram of the buck-converter.
\section{Embedded Software}
\begin{figure}[H] 
	\centering
\includegraphics[width=0.9\columnwidth]{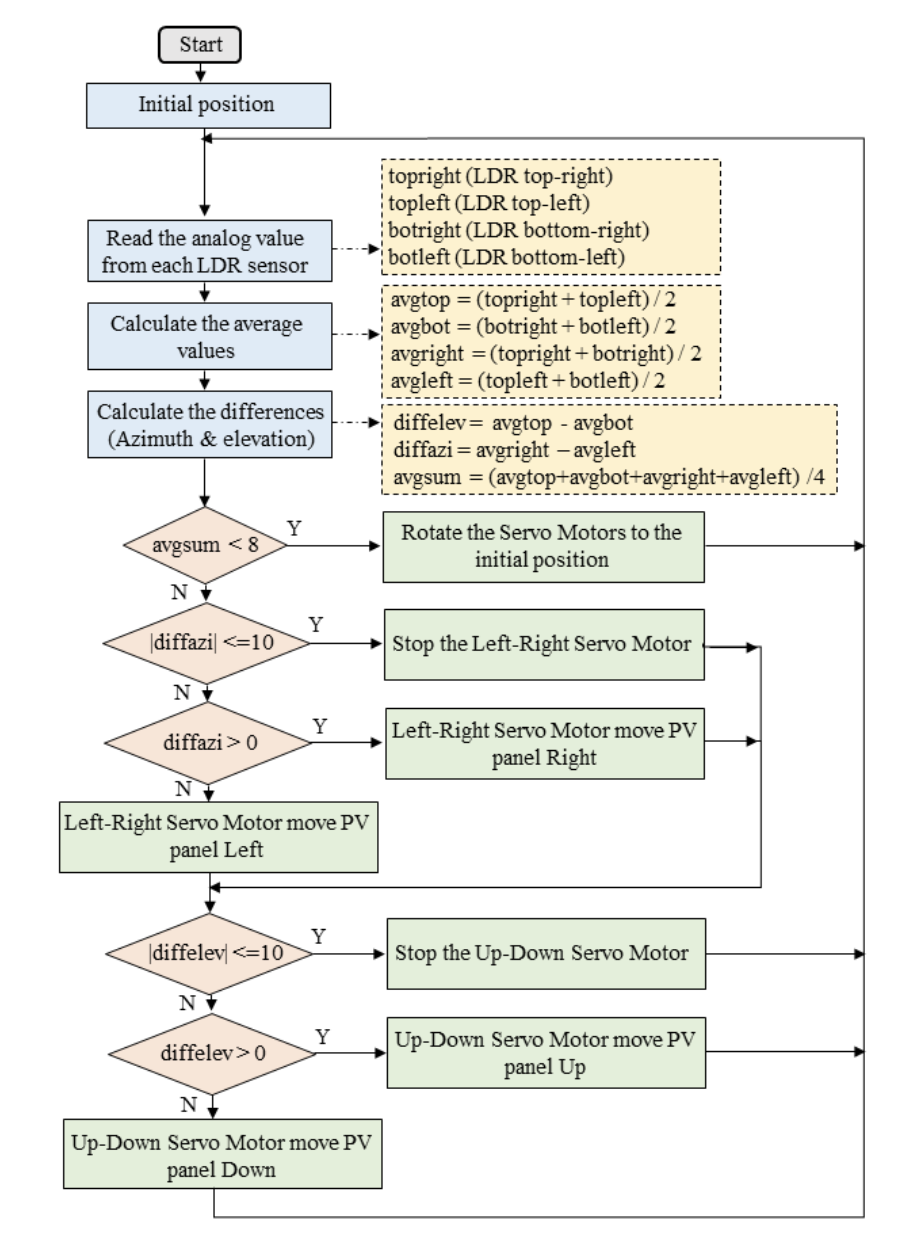} 
	\caption{Flowchart for Automatic Positioning of Solar Tracker.  Source: \cite{Elhammoumi:2020}}
	\label{fig:solar} 
\end{figure}

    Embedded code was written in C, compiled, and uploaded to the Arduino Rev Uno board using the Arduino IDE.  The model was simulated in Simulink using the MMPT algorithm.  The Arduino code for the embedded system control including soil moisture sensors, PWM timer to control stepper motor, and code to measure the water level using an ultrasonic sensor that was placed under the lid of the second water tank.   A solar tracking algorithm controls the angle change of the PV panel, and thus its rotation and tilt, based on the four light photocell resistor (LPR) measurements.  Figure \ref{fig:solar} provides a flowchart for an algorithm that automates the position mode of the solar tracker based on the measurement readings from the four LPR sensors.
    
    The Matlab code used for system simulation and embedded system control in Simulink included stepper motor control and step count. The Arduino and Matlab code is provided in the appendix.

\section{Results}
  
\subsection{Simulation}
\begin{figure}[H] 
	\centering  \includegraphics[width=0.9\columnwidth]{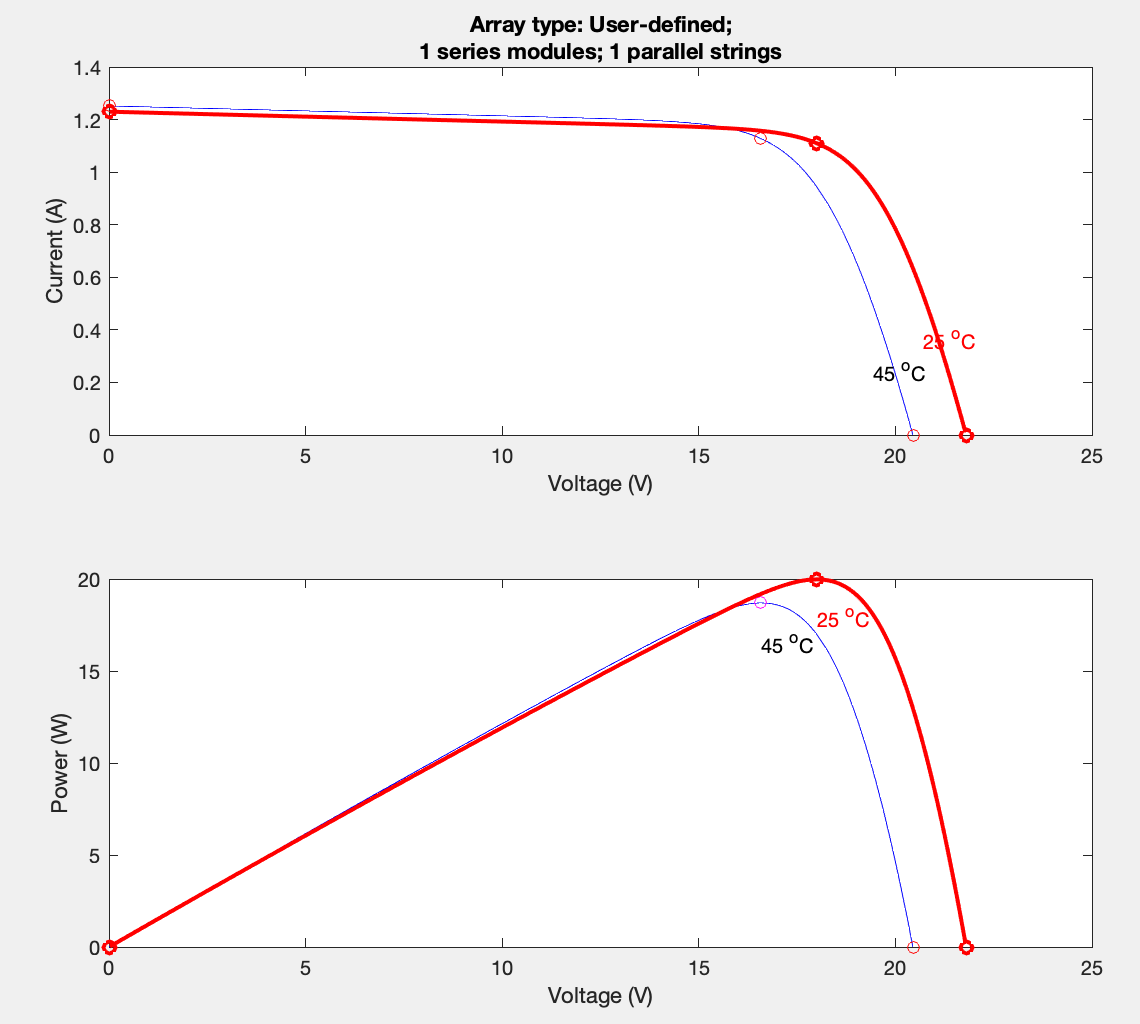} 
	\caption{IV and PV Performance of PV array @1000 $\text{W/m}^{2}$}
	\label{fig:model4} 
\end{figure}

\begin{figure}[H] 
	\centering  \includegraphics[width=0.9\columnwidth]{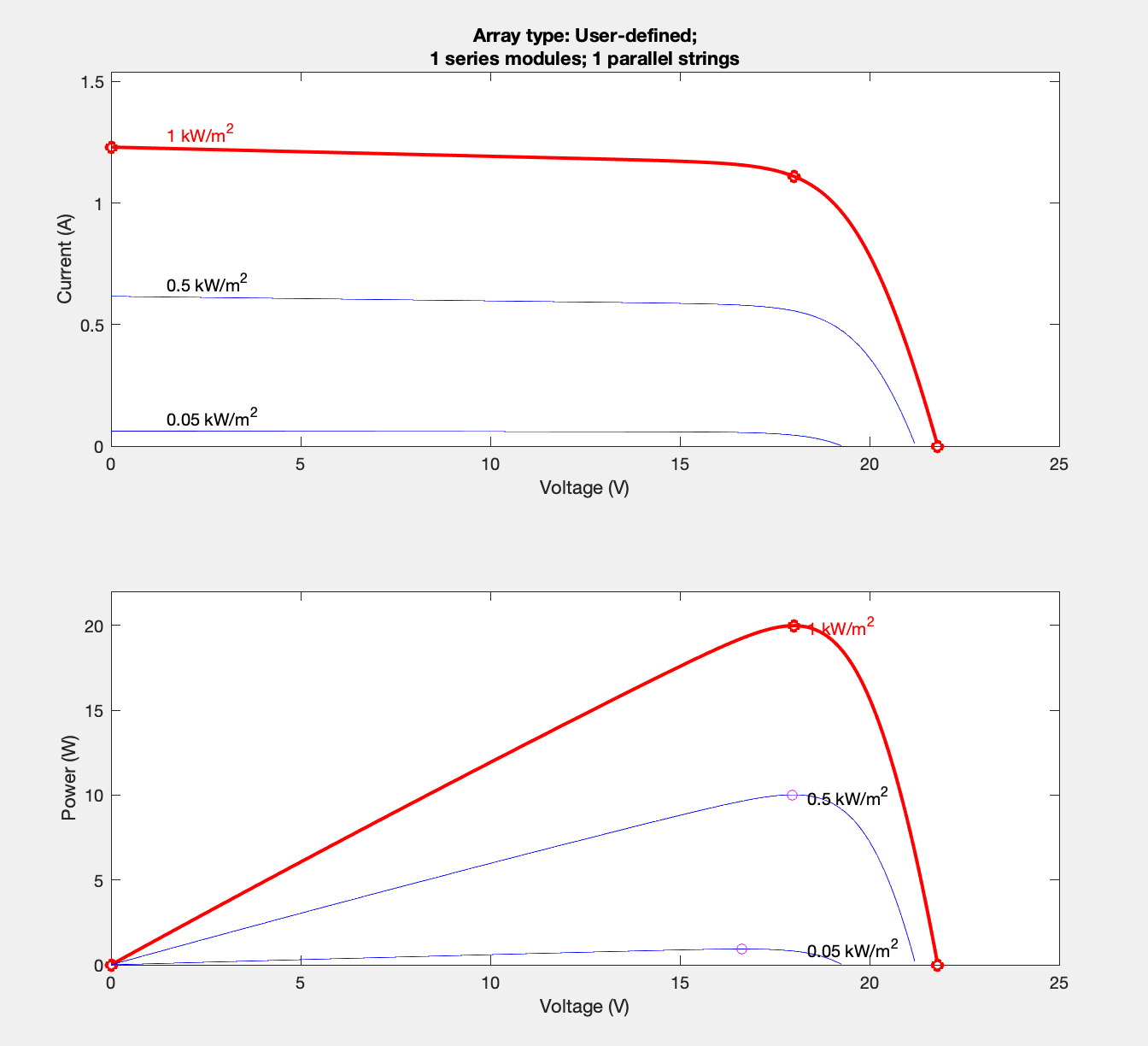} 
	\caption{IV and PV Performance @ $25^{\circ}$ C}
	\label{fig:model8} 
\end{figure}

    %Figure \ref{fig:model3} shows the Simulink diagram to generate (current-voltage) IV and PV (power-voltage) characteristics of a PV array @ $1000 W/m^{2}$.  
%\begin{figure}[H] 
%	\centering  \includegraphics[width=0.9\columnwidth]{Images/model3.png} 
%	\caption{Simulink Model to Generate IV-PV}
%	\label{fig:model3} 
%\end{figure} 
    Figure \ref{fig:model4} shows a plot of the IV and PV characteristics of the PV array @ $1000 W/m^{2}$. generated from the Simulink model MPPT algorithm assuming the maximum voltage of 200W.  As shown, the higher the temperature, the lower the power and current at voltages greater than 30.
     Figure \ref{fig:model8} shows plots of the PV array @ $25^{\circ}$ C and specified irradiances.  
    As the irradiance increases as measured in $\text{kW/m}^{2}$, the higher the maximum power point (MPP). 

\begin{figure}[H] 
	\centering  \includegraphics[width=0.9\columnwidth]{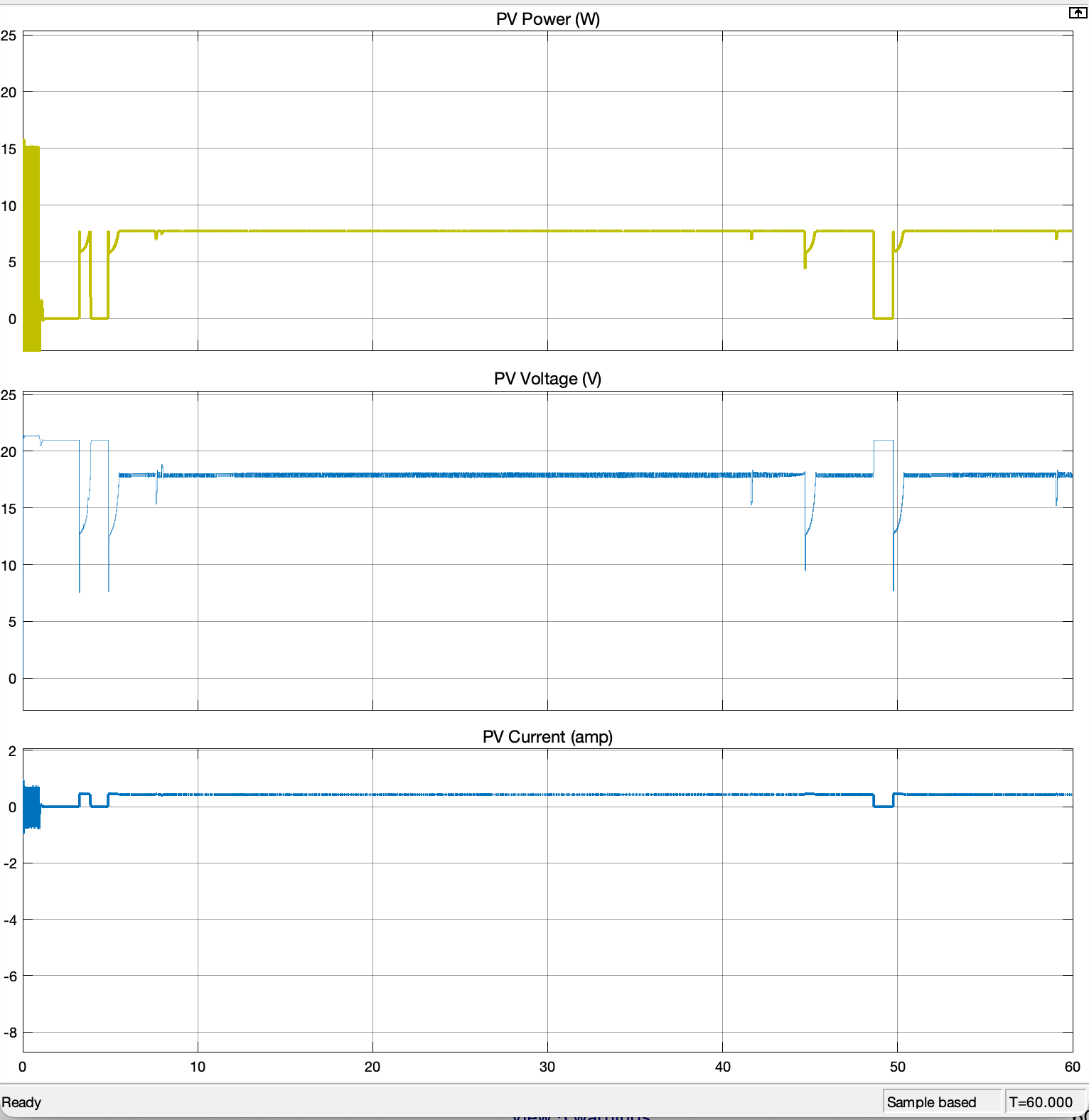} 
	\caption{Simulated PV voltage, current, and power}
	\label{fig:v2} 
    \end{figure}

\begin{figure}[H] 
	\centering  \includegraphics[width=0.9\columnwidth]{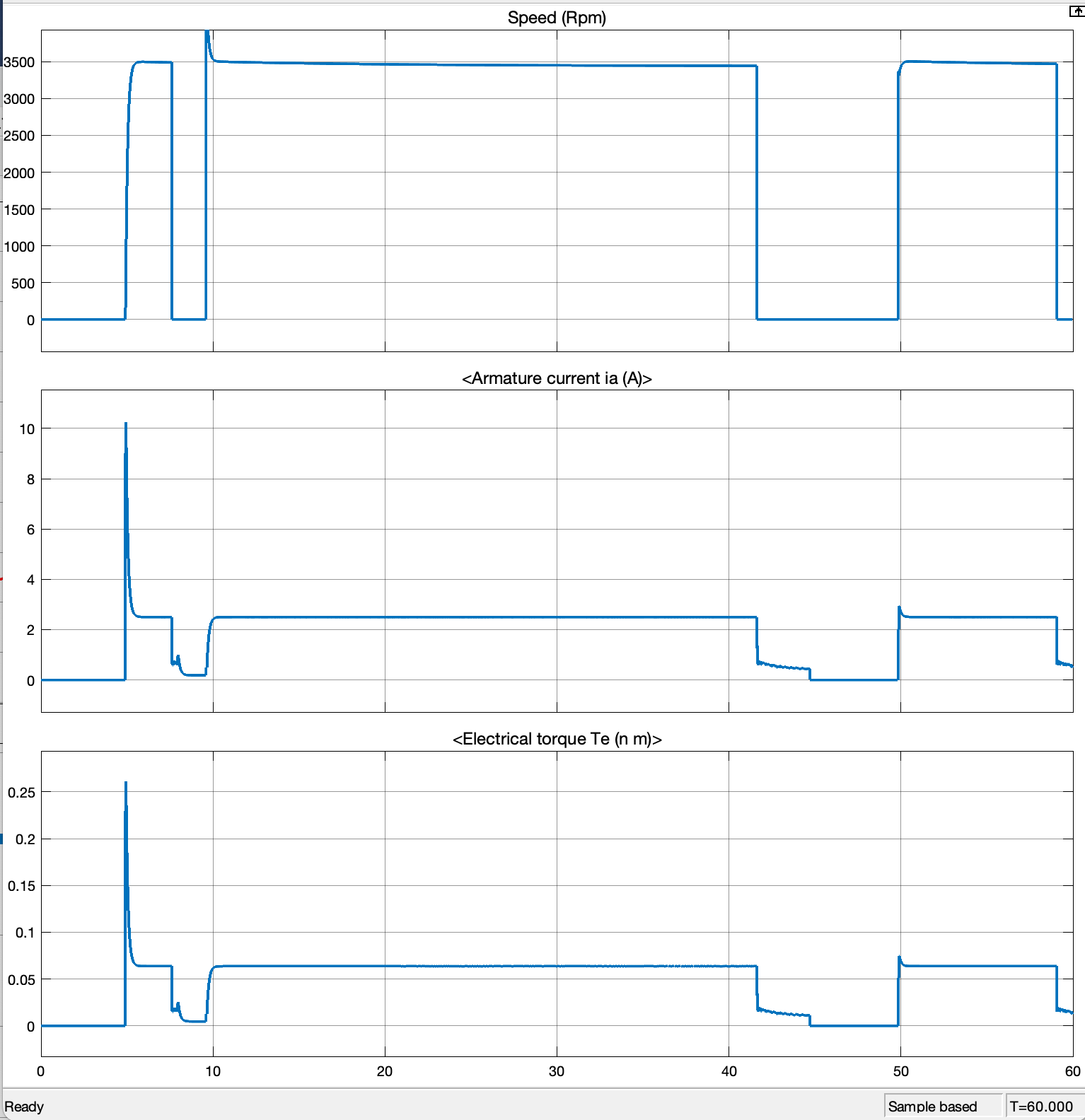} 
	\caption{Simulated motor speed, current, and torque}
	\label{fig:v4} 
    \end{figure}

    Figure \ref{fig:v2} shows the simulated voltage, current, and power generated by the PV panel.  In general, these values remain roughly constant over short periods of time since the MPPT algorithm matches the voltage of the PV to the  voltage of the battery that stores the converted solar energy as charge.
    Figure \ref{fig:v4} shows the stepper motor speed (Rpm), armature current (A), and electric torque (Nm) as the solar irradiance and light photocell resistor level changes. 

\begin{figure}[H] 
	\centering  \includegraphics[width=0.9\columnwidth]{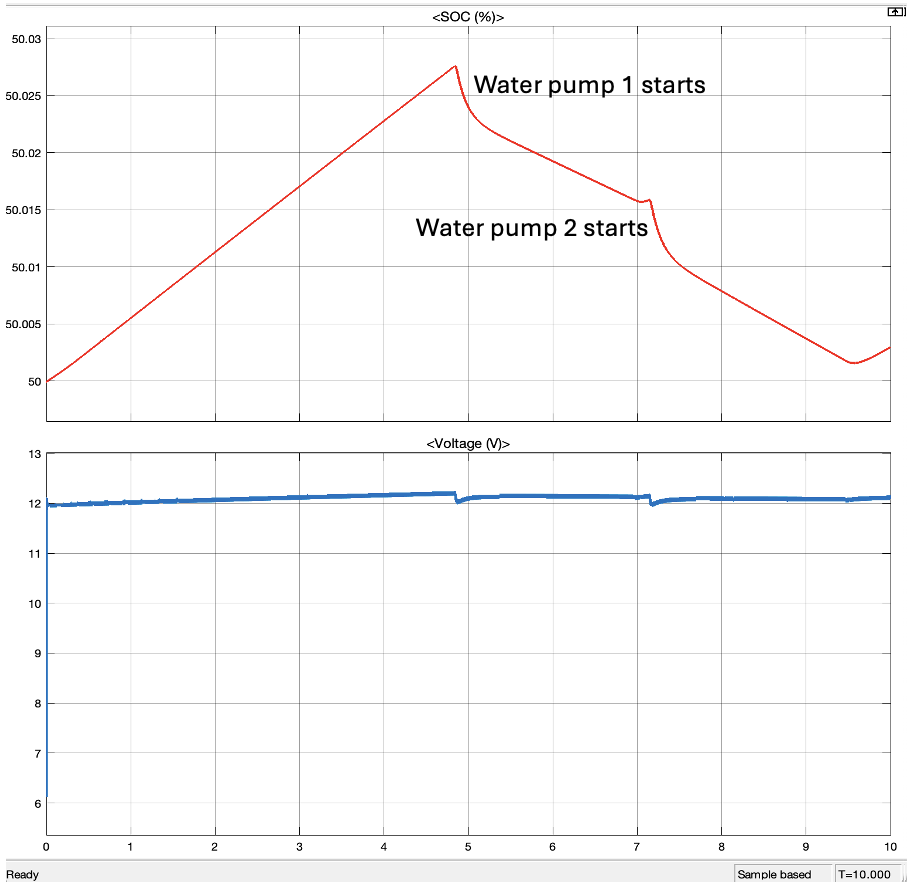} 
	\caption{SOC\% and voltage}
	\label{fig:current10} 
    \end{figure}

    Figure \ref{fig:current10} shows simulated battery state of charge (SOC) percentage of a cell.  The value at a given point in time denotes the capacity that is currently available as a function of the rated capacity.  
    As shown, initially the SOC\% increases as the battery is charged by the solar PV.  However, in the first SOC dissipation phase, water pump 1 start.  In the second dissipation phase, water pump 2 starts.  The voltage remains roughly constant except for small drops when the water pumps start as they serve as loads and draw power from the battery.
    
\begin{figure}[H] 
	\centering  \includegraphics[width=0.9\columnwidth]{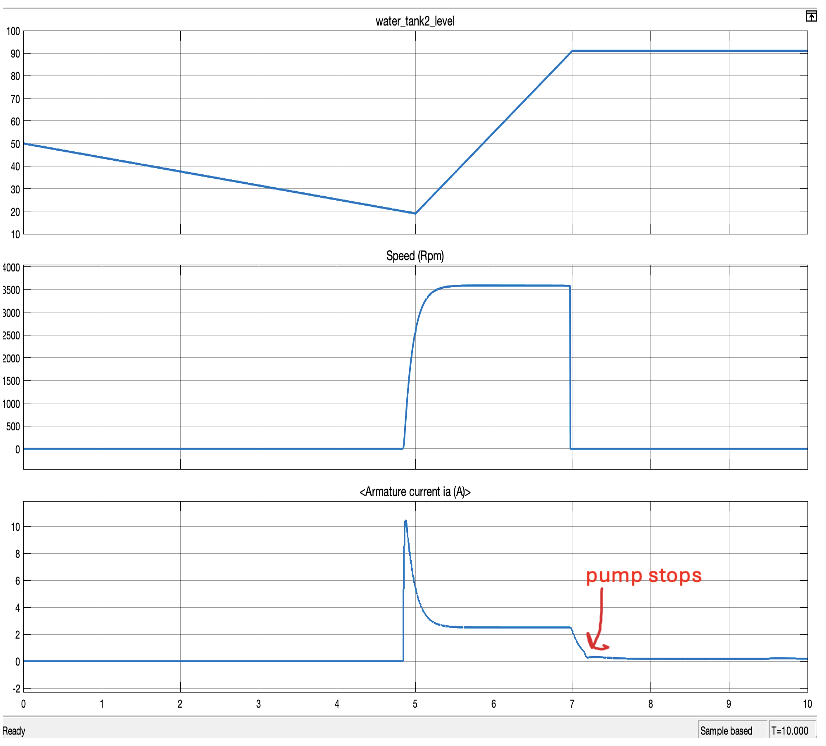} 
	\caption{Simulated water level of tank 2 and water pump 1 speed and armature}
	\label{fig:water} 
    \end{figure}

\begin{figure}[H] 
	\centering  \includegraphics[width=0.9\columnwidth]{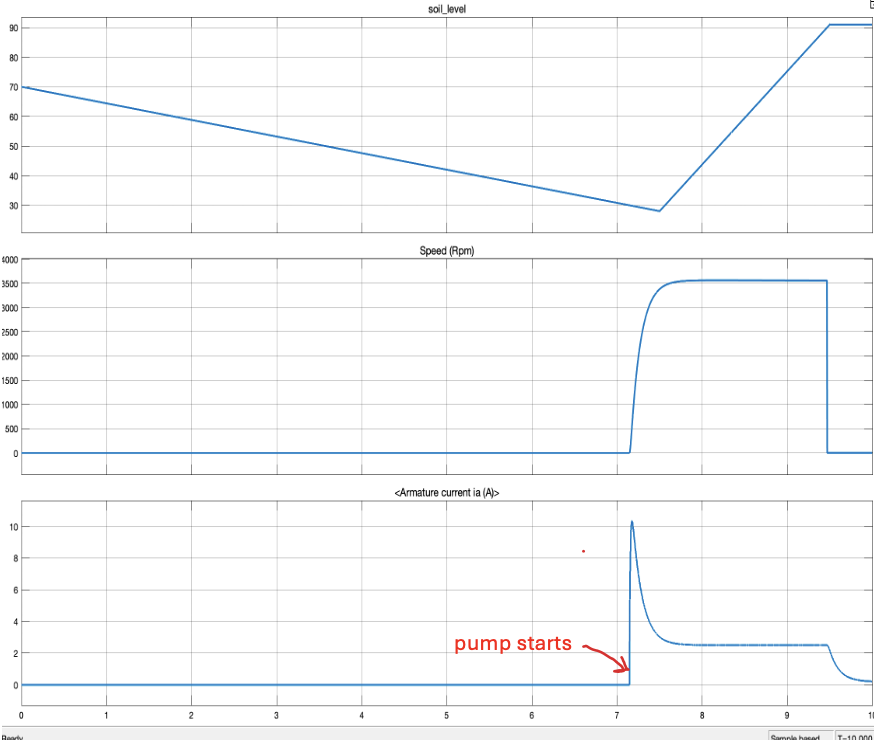} 
	\caption{Simulated soil moisture level}
	\label{fig:water2} 
    \end{figure}

\begin{figure}[H] 
	\centering  \includegraphics[width=0.9\columnwidth]{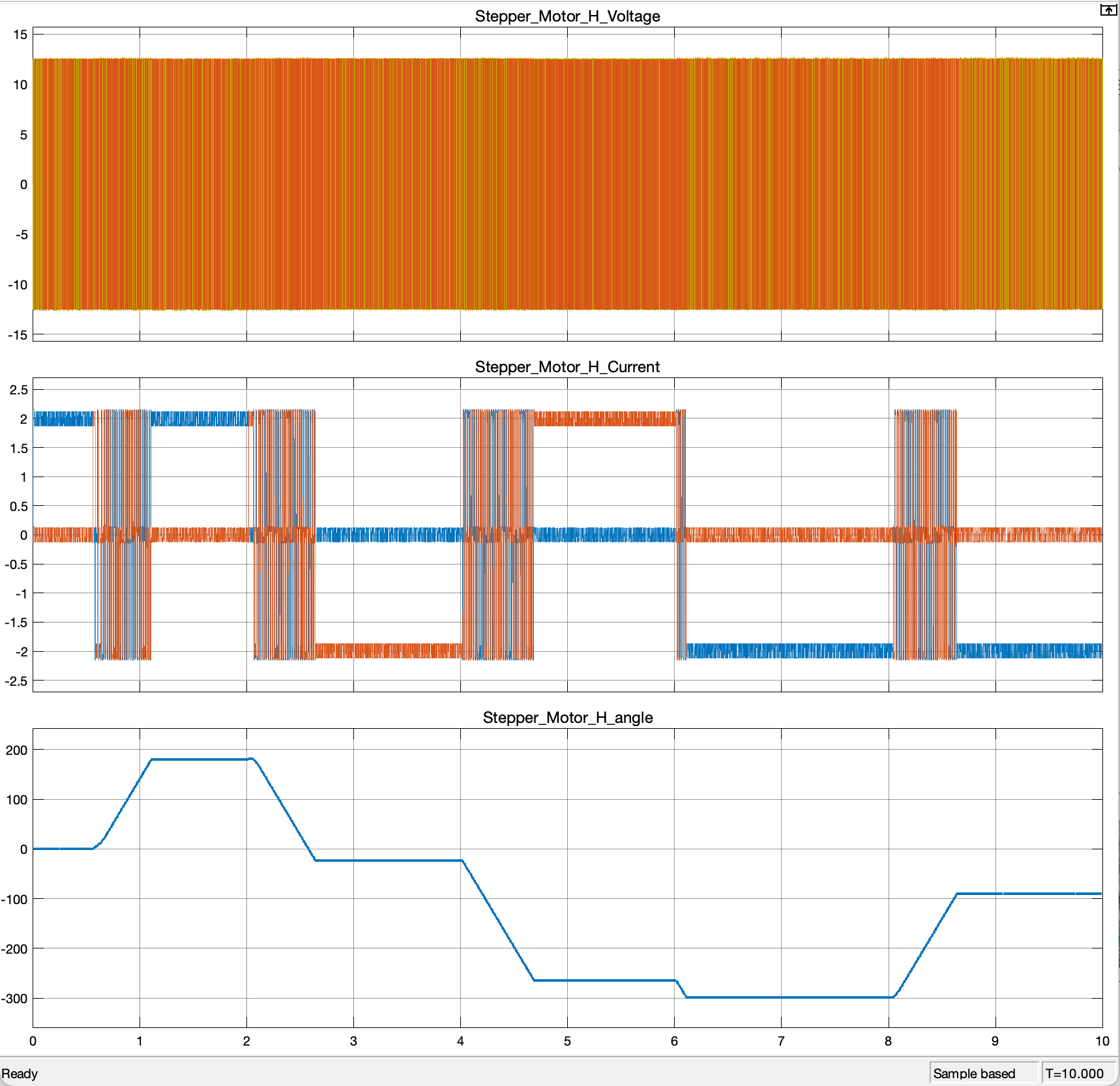} 
	\caption{Stepper motor H voltage, current, and angle}
	\label{fig:current} 
    \end{figure}

\begin{figure}[H] 
	\centering  \includegraphics[width=0.9\columnwidth]{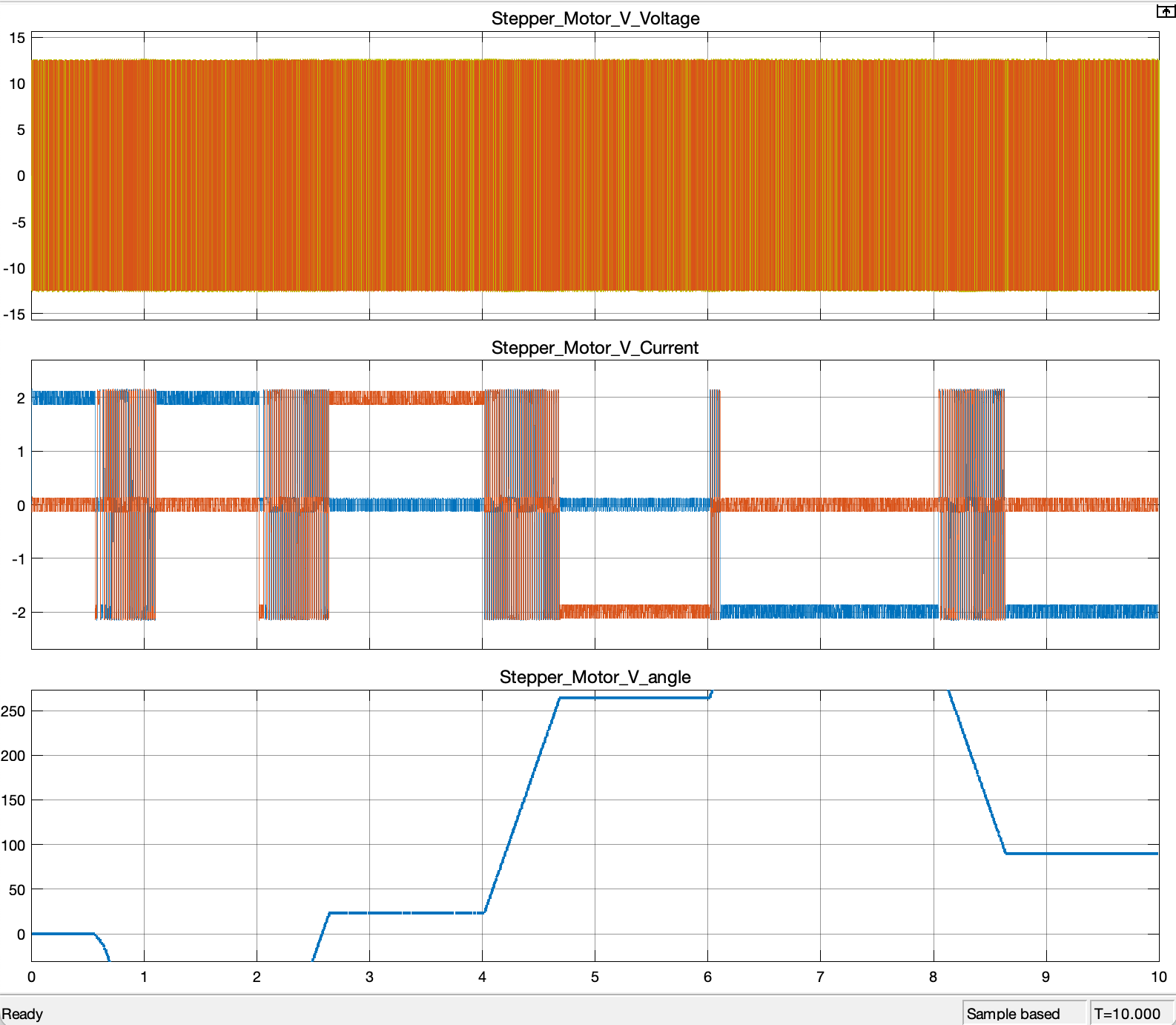} 
	\caption{Stepper motor H voltage, current, and angle}
	\label{fig:current2} 
    \end{figure}

    Figure \ref{fig:water} shows the water tank level in tank 2.  When the water level fall below 20\%, water pump 1 pumps water to tank 2 which until it is full.  Both the speed and armature current of water pump 1 increase until the relay switch turns it off once the water level sensed by the ultrasonic sensor sense the tank is full at 90\%.
Figure \ref{fig:water} shows the stepper motor speed (Rpm), armature current (A), and electric torque (Nm) as the solar irradiance and light photocell resistor level changes. 
When the soil moisture level falls below a humid threshold, e.g., at a dry level, then water pump 2 turns on and pumps water through the hose and nozzle spray to the plant until the soil moisture level is wet.  Figure \ref{fig:water2} simulates the soil moisture level, water pump 2 speed  (rpm), and the pump's armature
Figure \ref{fig:current} and Figure \ref{fig:current2} show simulated horizontal stepper motor and vertical stepper motor voltage, current, and angle, respectively.

\begin{figure}[H] 
	\centering  \includegraphics[width=0.9\columnwidth]{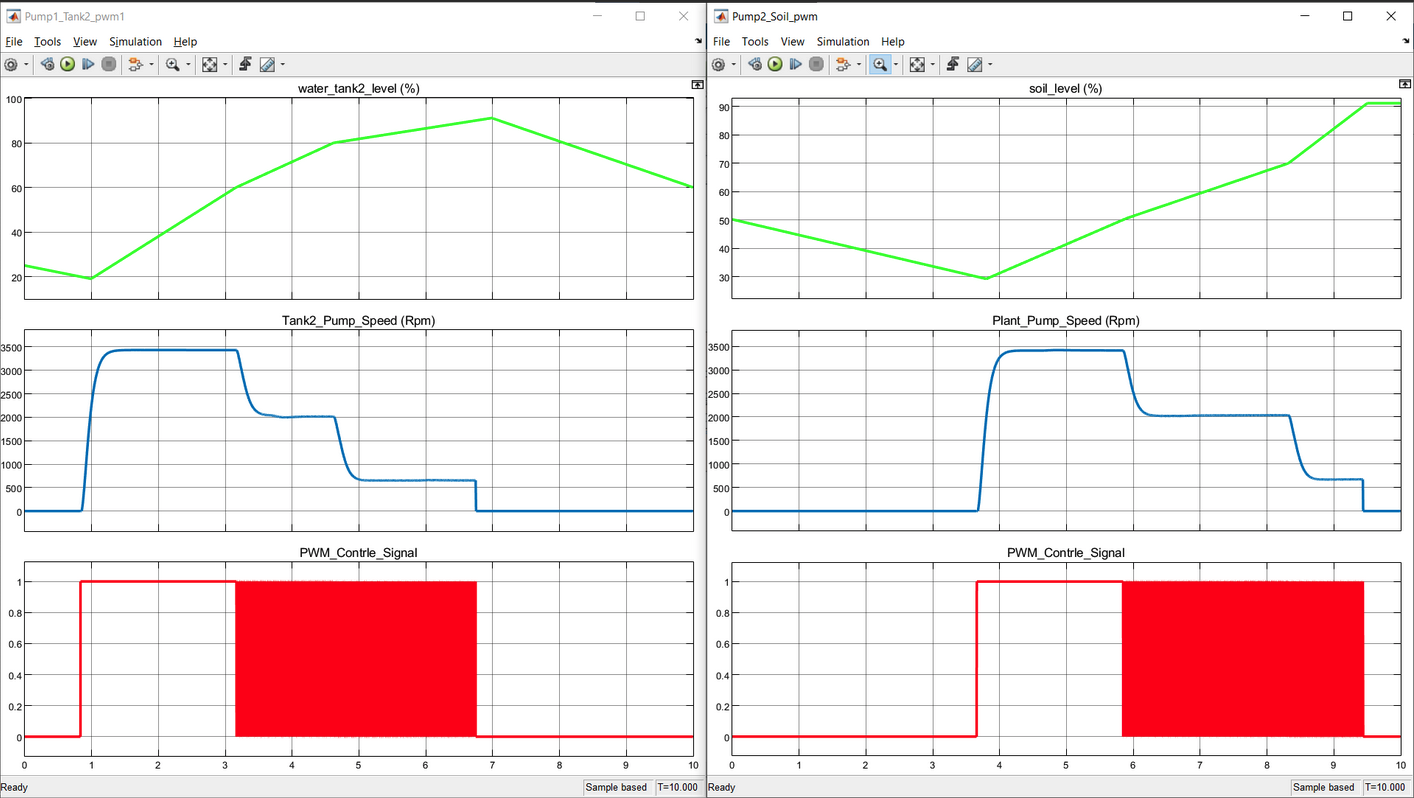} 
	\caption{PMW control signals generated for water tank 2 levels (left) and soil level percentage (right)}
	\label{fig:pwm} 
\end{figure}

Figure \ref{fig:pwm} shows simulated PWM control signals based on the water tank 2 level (left) and the soil level percentage (right).  As shown, water pump 1 turns on when the water tank 2 falls below 20\%.  At this time, the PWM signal jumps to 1 and the motor speed of water tank 1 increases from 0 to just under 3500 rpm.  

\begin{figure}[H] 
	\centering  \includegraphics[width=0.9\columnwidth]{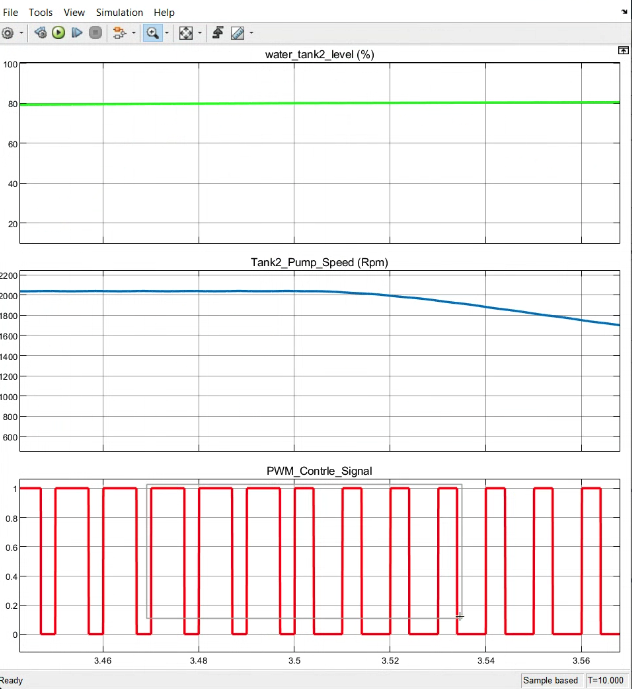} 
	\caption{Rectangular pulses of PWM signal (zoomed in)}
	\label{fig:pwm7} 
\end{figure}

After tank 2 is 60\% full, the water pump speed slows down and the PMW control signal fluctuates rapidly as rectangular pulses between 0 (\enquote{off}) and 1 (\enquote{on}) as it is approximating a continuous changing pump speed analog signal. The duty cycle is approximately 50\% during this time.   When the water level in tank 2 reaches 85\%, water pump 1 turns off and PWM signal becomes 0. 
Figure \ref{fig:pwm7} shows the PWM signal zoomed in when the soil moisture percentage increases above 60\% making it easier to see the rectangular pulses.
    
Likewise, when the soil moisture percent falls below 30\%, the motor pump speed of water pump 2 increases causing a jump in the PWM signal as water is pumped from water tank 2 to the plant.  When the soil moisture hits 50\%, the speed of the water pump slows down until it reaches 90\% when water pump 2 shuts off via a relay switch.  The PMW signal also fluctuates rapidly between 0 and 1 until the pump is turned off.

\subsection{Experimental}
\begin{figure}[H] 
	\centering  
     \includegraphics[width=0.9\columnwidth]{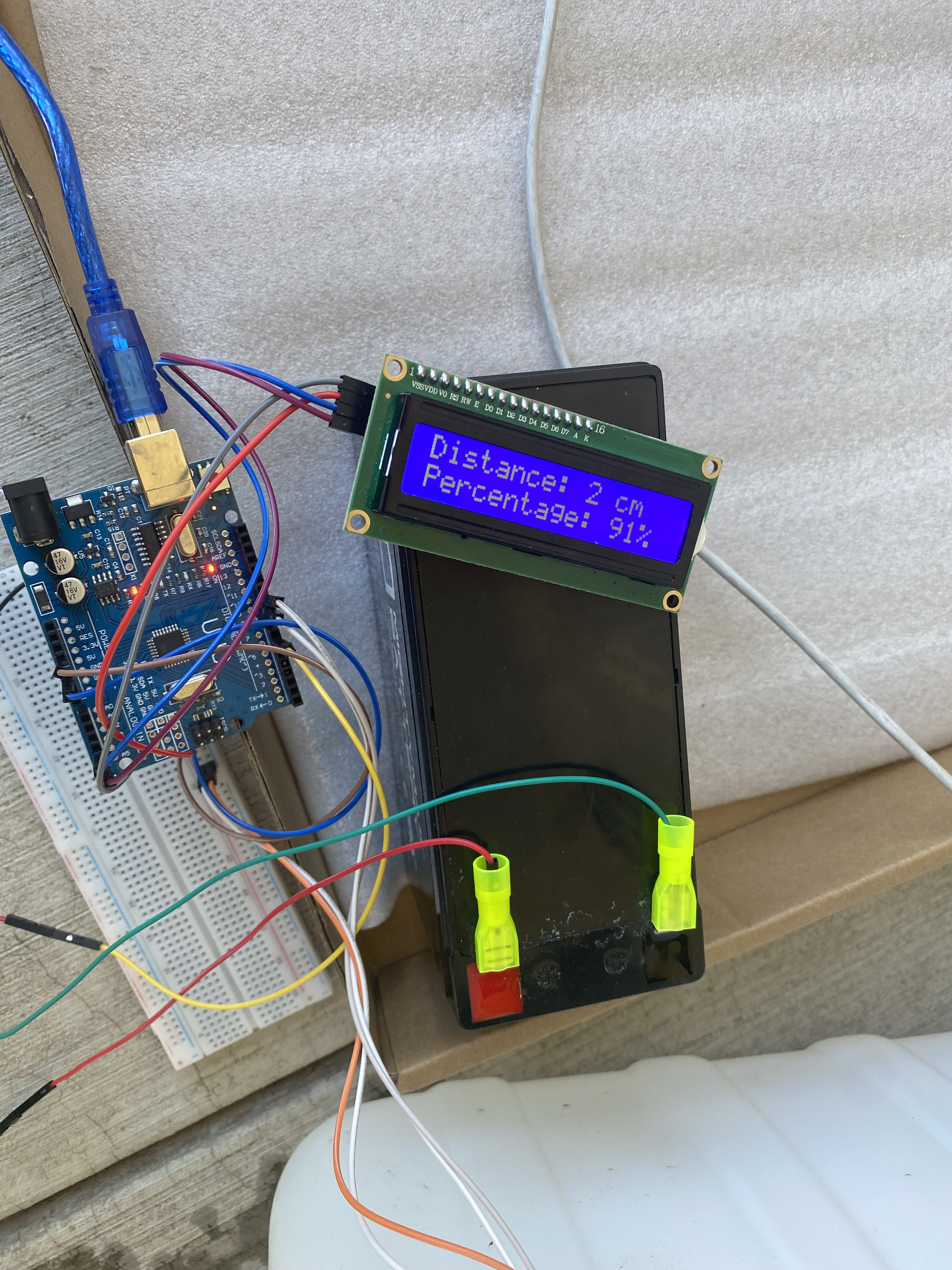} 
	\caption{}
	\label{fig:bread} 
\end{figure}

\begin{figure}[H] 
	\centering  
     \includegraphics[width=1\columnwidth]{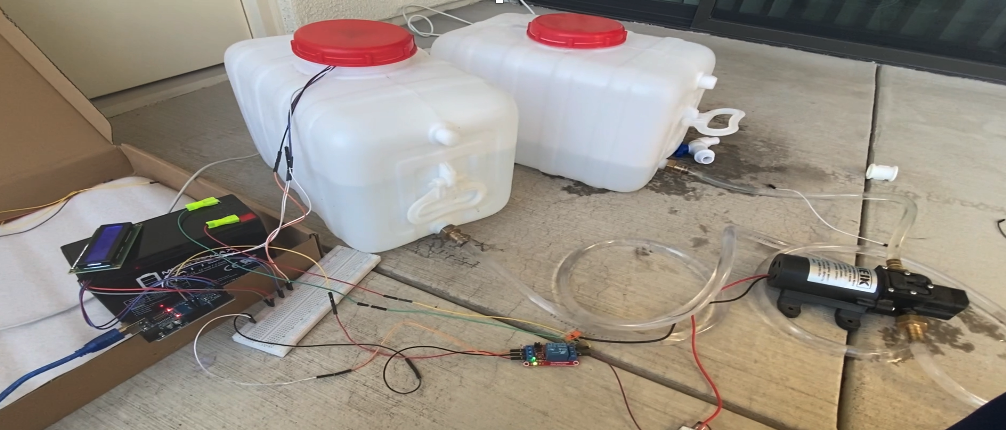} 
	\caption{Arduino and Breadboard}
	\label{fig:setup1} 
\end{figure}

\begin{figure}[H] 
	\centering  
     \includegraphics[width=0.9\columnwidth]{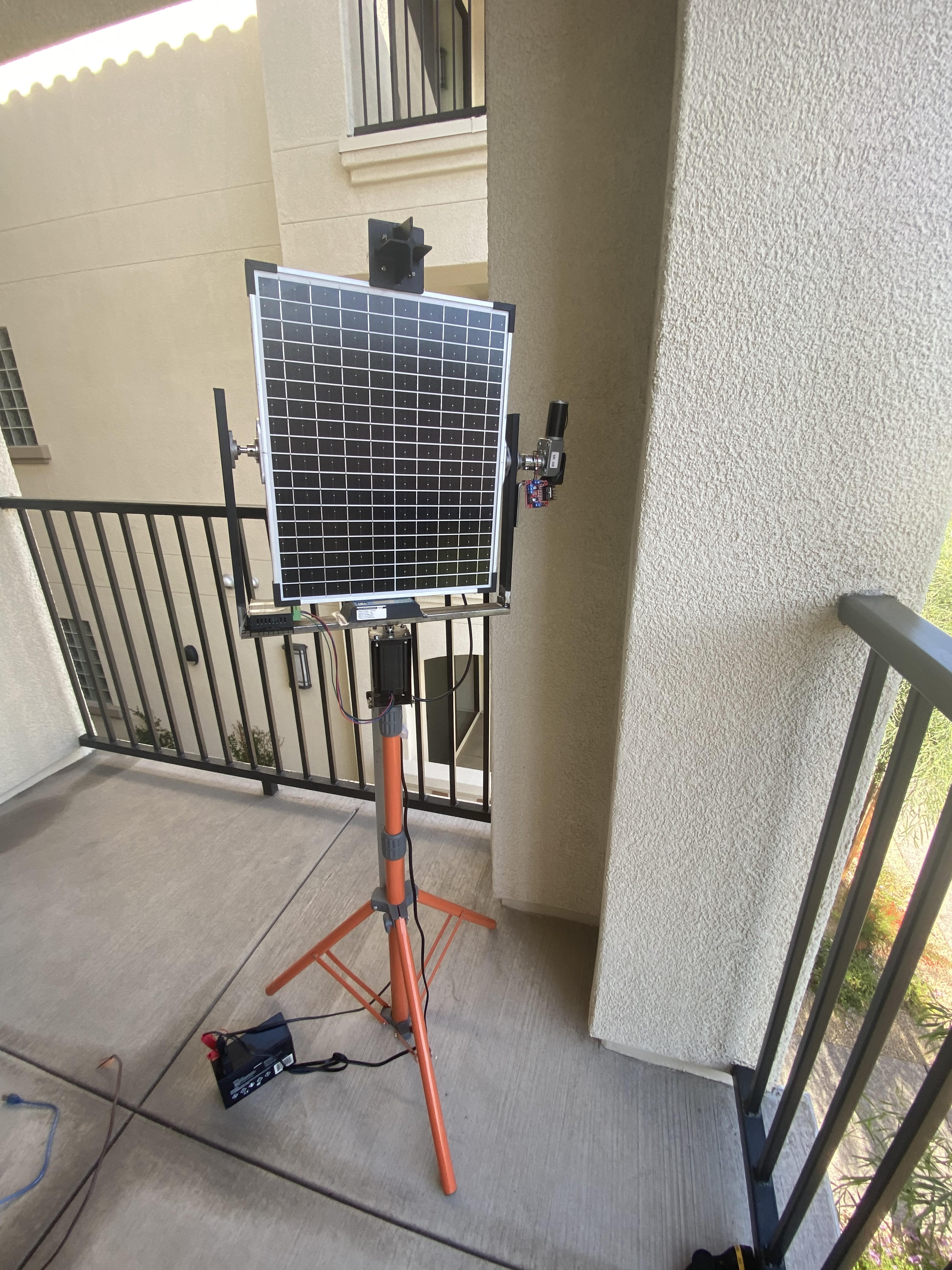} 
	\caption{Solar Tracker on tripod}
	\label{fig:setup2} 
\end{figure}

\begin{figure}[H] 
	\centering  
     \includegraphics[width=0.9\columnwidth]{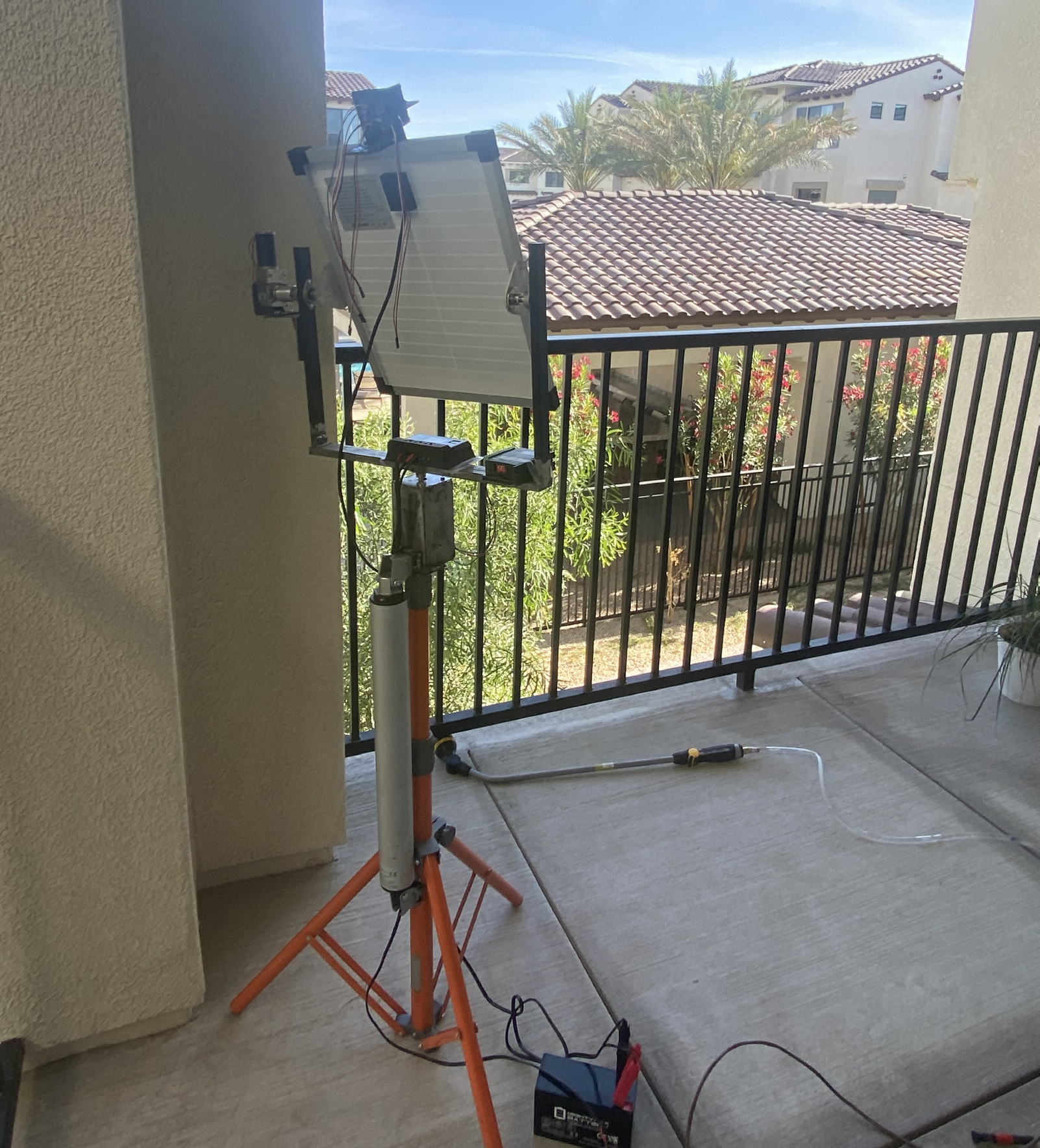} 
	\caption{Solar Tracker facing sunlight}
	\label{fig:solar7} 
\end{figure}

\ \ \ Figure \ref{fig:bread} shows a display of the water level measured by the ultrasonic sensor and shown on the LCD display along with wiring to the Arduino and breadboard.  The ultrasonic sensor measures the distance from the top of the tank to the top of the water.  As the water level rises in tank 2 from pumped water from tank 1, the total water percentage in the tank increases and the ultrasonic sensor measured distance decreases.
Figure \ref{fig:setup1} shows the two water tanks with the ultrasonic sensor wires coming out from underneath the lid of the second water tank along with wiring to the Arduino and breadboard.
Figure \ref{fig:setup2} shows the DC motor (on right arm) and stepper motor (at base) connected to the PV panel on the tripod.
Figure \ref{fig:solar7} shows a different angle of solar tracker with the front side of the panel facing the sunlight.
Figure \ref{fig:light} shows the four LPRs cells in 4 quadrants.  The tracking algorithm determines the optimal rotation and tilt adjustment of the  PV panel based on the relative intensities between the LRPs.  For instance, if the intensity is greatest in the northeast (first quadrant), the stepper and DC motors will move the panel to optimize the sunlight in this direction. 
\begin{figure}[H] 
	\centering  
     \includegraphics[width=0.9\columnwidth]{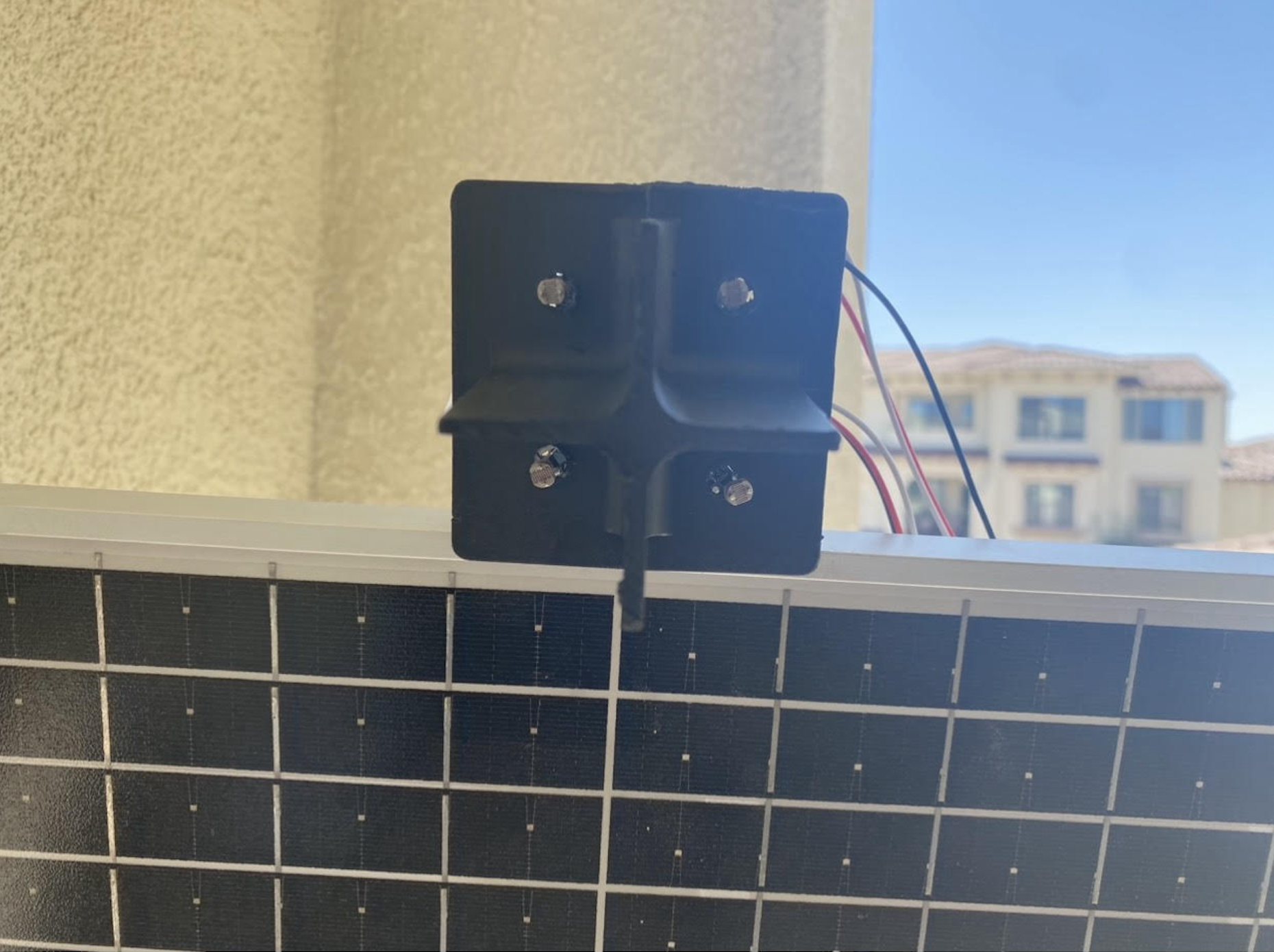} 
	\caption{LPR in 4 quadrants}
	\label{fig:light} 
\end{figure} 
Figure \ref{fig:LPR} shows actual LPR values generated by each of the LPRs when the temperature outside was roughly $81^{\circ}$C with sunny skies. As shown, most values stay relatively constant until the solar irradiance changes based on the position of the sun and/or the temperature changes.
\begin{figure}[H] 
	\centering  
     \includegraphics[width=0.9\columnwidth]{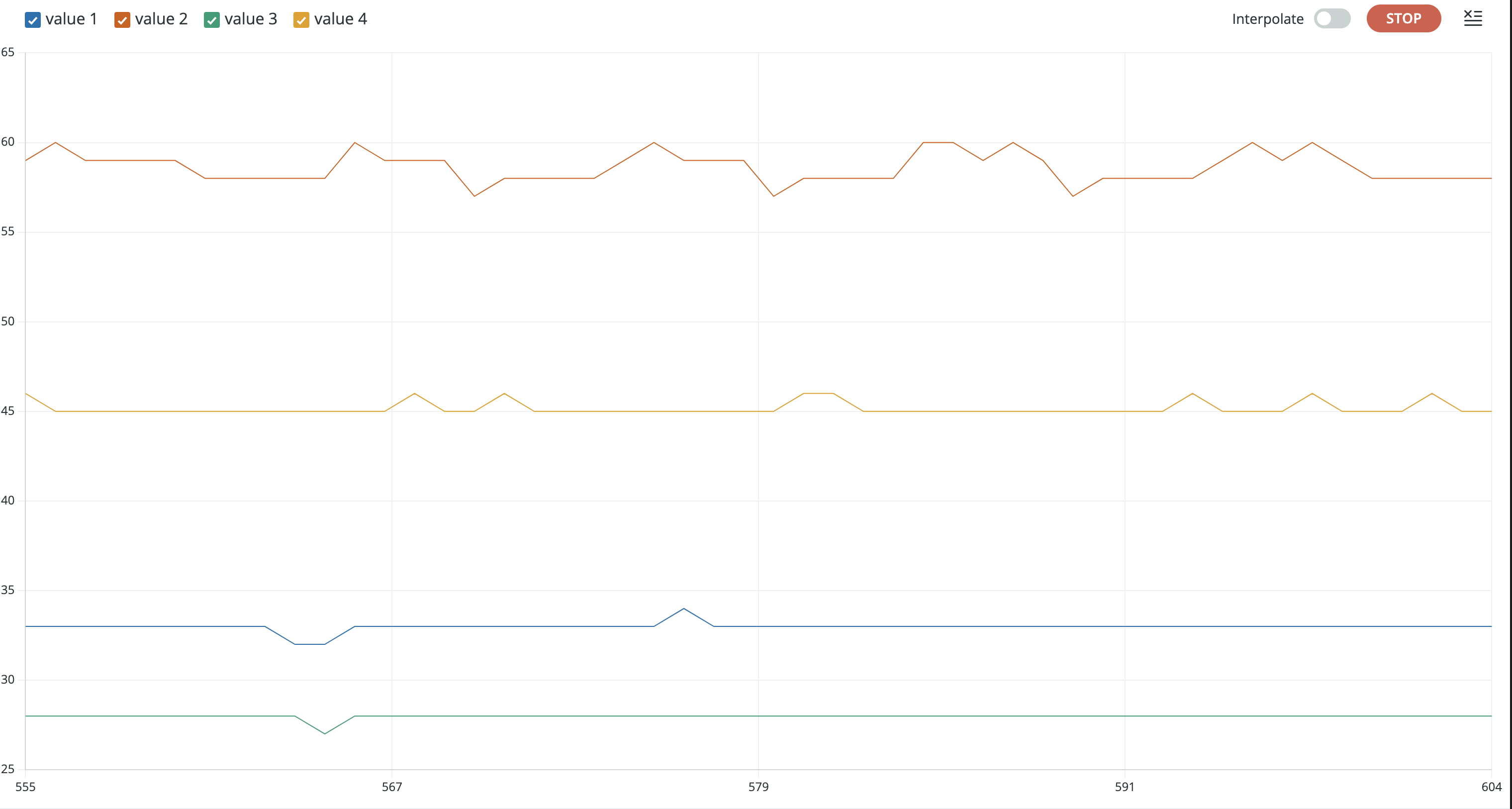} 
	\caption{LPR Values}
	\label{fig:light} 
\end{figure} 
Figure \ref{fig:setup4} shows the solar charge controller measuring the charge to the 12 V lead acid battery and the TB6600 stepper motor driver.
\begin{figure}[H]
	\centering  
     \includegraphics[width=0.9\columnwidth]{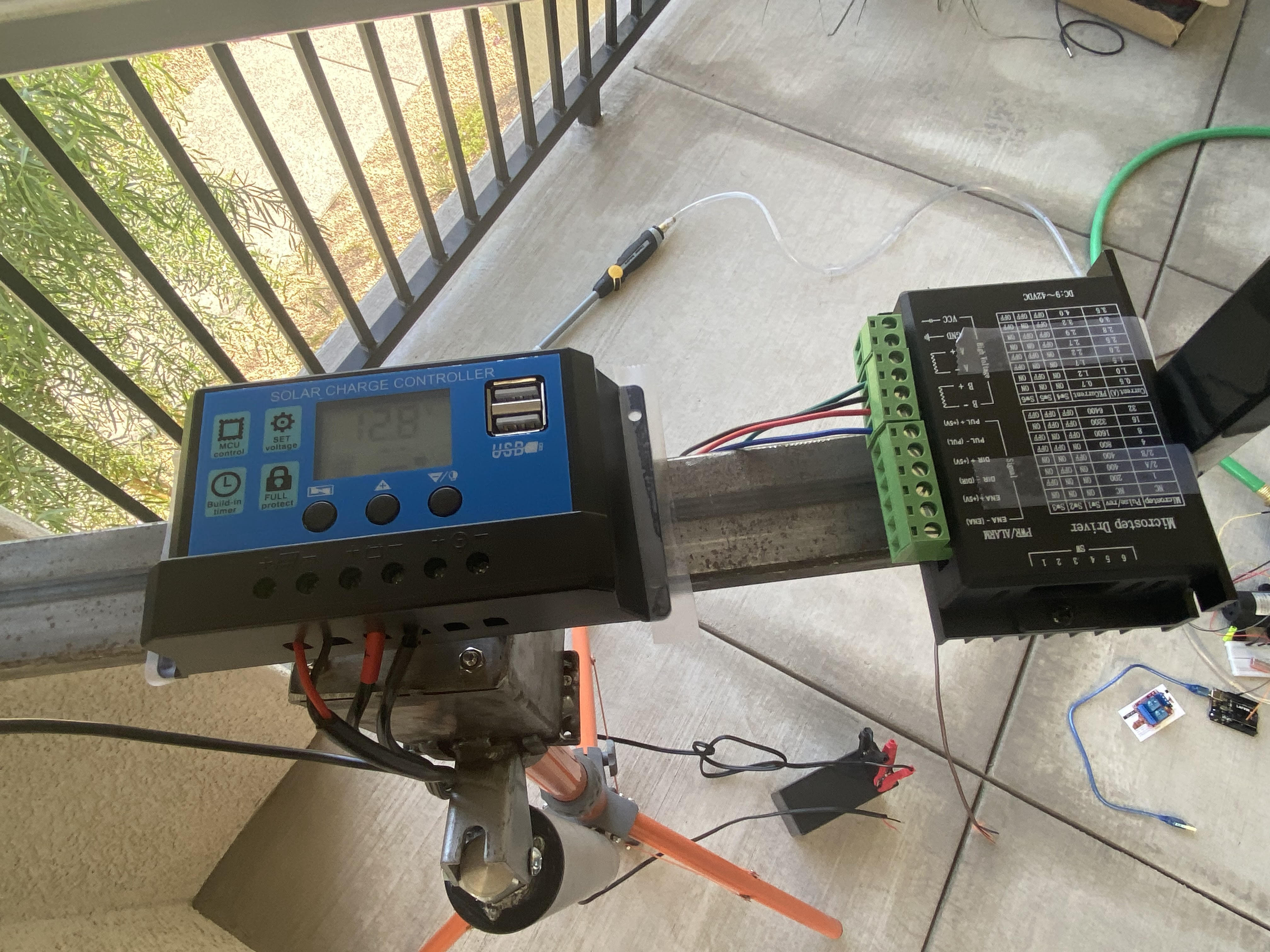} 
	\caption{Solar charge control}
	\label{fig:setup4} 
\end{figure}
    Figure \ref{fig:setup4} shows the solar tracker raised up by the linear actuator.
\begin{figure}[H]
	\centering  
     \includegraphics[width=0.75\columnwidth]{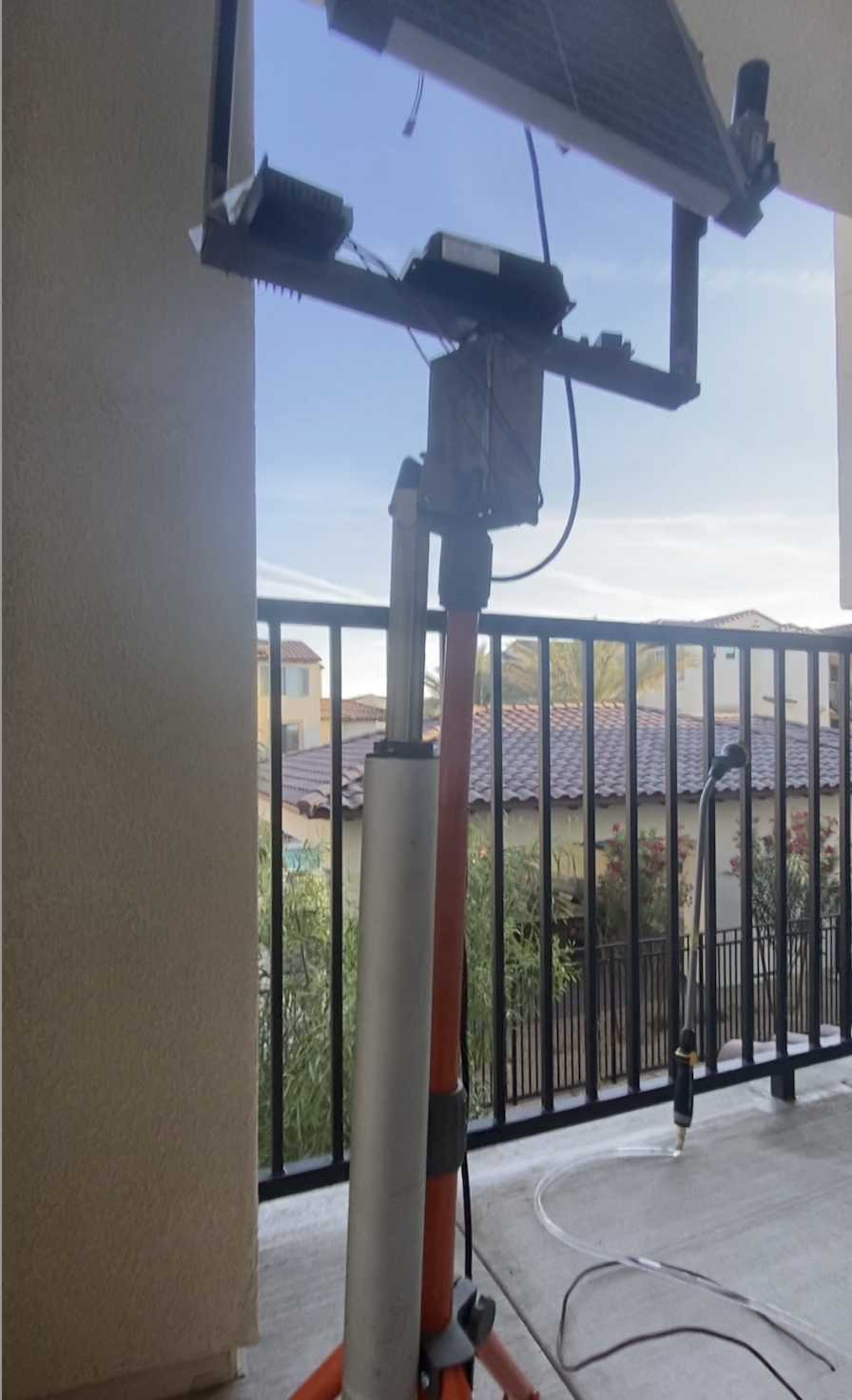} 
	\caption{Linear actuator}
	\label{fig:setup4} 
\end{figure}

\section{Summary}
    A novel 3-axis solar tracker was introduced that could rotate and tilt using stepper motors and rise up and down on tripod with a linear actuator.  The results show that the proposed system is efficient and cost-effective as a method for watering plants, horticulture, or irrigation in rural regions that do not have access to national power grid distribution. 
    
    An Arduino board was used as an embedded microcontroller with embedded software to measure water level and light intensity by communicating with an ultrasonic sensor and 4 LPR sensors attached to the PV, respectively, via I2C/SPI.  The embedded C code controlled the flow of water from tank 1 to tank 2 by turning relay switches on and off thereby opening and closing the motorized ball valves. 

    The simulated results illustrate power-voltage (PV) and current-voltage (IP) characteristics of the PV panel, how PWM can control the speed of the water pumps, and how the motor speed, current, and torque of the stepper motors changes with LPR sensor measurements of solar irradiance.
    
    The experimental results illustrate how the solar charge controller and Arduino board MCU can be used with embedded software and algorithms to control the switching on and off of water pumps based on sensor measurements.  For instance, The MPPT perturbation and observation algorithm serves as a DC-DC converter to ensure the voltage of the 20W PV panel matches the voltage of the 12V lead acid battery as solar energy is stored as  electrical charge.

    Despite the promising results, there are various challenges that need to be resolved for the prototype system to be modified for use in a commercial application.  First, the various number of crossing jumper wires, motor drivers, relay switches, breadboards, and wago connectors make diagnosing power issues difficult even with a voltmeter.  Second, the stepper motor requires quite a bit of tuning and configuration via changing the TB6600 motor driver DIP switches to get the proper microstepping.  Third, the solar tracker frame weight impacts the microstepping performance because the shaft requires higher torque to rotate both the frame and PV.
    
    Future work will involve designing a PCB with integrated circuits (ICs) and frame fabrication using lighter materials such as aluminum instead of steel to reduce the torque needed for microstepping.  At the same time,  the frame weight cannot be too low as it must be robust to wind gusts.

%\begin{figure}[H] 
%	\centering
%\includegraphics[width=0.9\columnwidth]{Images/code1.png} 
%	\caption{Stepper motor control}
%	\label{fig:control1}  
%\end{figure}
\newpage
\printbibliography

\end{document}